\documentclass[conference]{IEEEtran}
\usepackage{float}
\usepackage{subfig}
\usepackage{tabularx}
\usepackage{amsmath}
\usepackage{graphicx}
\usepackage{amssymb}
\graphicspath{{./figures/ }}
\usepackage[rightcaption]{sidecap}
\usepackage{wrapfig}
\usepackage{xcolor}
\usepackage{multirow}
\usepackage{url}
\usepackage{comment}
\usepackage{pifont}
\pdfoutput=1

\newcommand{\cmark}{\ding{51}}%
\newcommand{\xmark}{\ding{55}}%
\definecolor{gatororange}{RGB}{250,70,22}
\definecolor{wow}{RGB}{50,10,122}
\definecolor{muse}{RGB}{225,23,133}
\definecolor{blue}{RGB}{0,0,255}

\newcommand{\ignore}[1]{\if{0} #1 \fi}
\newcounter{inlineenum}
\renewcommand{\theinlineenum}{\alph{inlineenum}}
\newenvironment{inlineenum}
  {\unskip\ignorespaces\setcounter{inlineenum}{0}%
   \renewcommand{\item}{\refstepcounter{inlineenum}{\textit{\theinlineenum})~}}}
  {\ignorespacesafterend}
\newcounter{subsubsubsection}[subsubsection]

\begin{document}
\title{Hear ``No Evil'', See ``Kenansville''*: Efficient and Transferable
Black-Box Attacks on Speech Recognition and Voice Identification Systems}

\author{\IEEEauthorblockN{Hadi Abdullah, Muhammad Sajidur Rahman, Washington Garcia, Logan Blue,\\ Kevin Warren, Anurag Swarnim Yadav, Tom Shrimpton and Patrick Traynor}
\IEEEauthorblockA{University of Florida\\
\{hadi10102, rahmanm, w.garcia, bluel, kwarren9413, anuragswar.yadav, teshrim, traynor\}@ufl.edu
}
}

\maketitle
{\let\thefootnote\relax\footnotetext{*The title of our paper plays on
``Hear No Evil, See No Evil'' and we use the attacks described in
our paper to generate the above title. Thus, when a model is fed ``No Evil'',
it mistranscribes it as ``Kenansville'', a town located in Central
Florida - text completely unrelated to the audio input.}}

\begin{abstract}

Automatic speech recognition and voice identification systems are being deployed in a wide array of applications,
from providing control mechanisms to devices lacking traditional interfaces, to the
automatic transcription of conversations and authentication of users.
Many of these applications have significant security and privacy considerations.
%
We develop attacks that force mistranscription and
misidentification in state of the art systems, with minimal impact on
human comprehension.  Processing pipelines for modern systems are comprised of signal
preprocessing and feature extraction steps, whose output is fed to a
machine-learned model.   Prior work has focused on the models,
using white-box knowledge to tailor model-specific attacks.  We focus on the
pipeline stages before the models, which (unlike the models) are quite
similar across systems.  As such, our attacks are black-box and
transferable, and demonstrably achieve mistranscription and misidentification rates as high as
100\% by modifying only a few frames of audio. 
We perform a study via
Amazon Mechanical Turk demonstrating that there is no statistically significant
difference between human perception of regular and perturbed audio. 
Our findings suggest that models may learn aspects of speech that are generally
not perceived by human subjects, but that are crucial for model accuracy. We also
find that certain English language phonemes (in particular, vowels) are
significantly more susceptible to our attack.  We show that the
attacks are effective when mounted over cellular networks, where
signals are subject to degradation due to transcoding, jitter, and
packet loss.

\end{abstract}

\section{Introduction}

The telephony network is still the most widely used mode of audio communication on the planet, with billions of phone calls occurring every day within the USA alone~\cite{usa_phone_calls}. Such a degree of activity makes the telephony network a prime target for mass surveillance by governments. However, hiring individual human listeners to monitor these conversations can not scale. To overcome this bottleneck, governments have used Machine Learning (ML) based Automatic Speech Recognition (ASR) systems and Automatic Voice Identification (AVI) systems to conduct mass surveillance of their populations. Governments accomplish this by employing ASR systems to flag anti-state phone conversations and AVI systems to identify the participants~\cite{inside-chinas-massive-surveillance-operation}. The ASR systems convert the phone call audio into text. Next, the government can use keywords searches on the audio transcripts to flag potentially dissenting audio conversations~\cite{nsa-speech-recognition-snowden-searchable-text}. Similarly, AVI systems identify the participants of the phone call using voice signatures. 

Currently, there does not exist any countermeasure for a dissident attempting to circumvent this mass surveillance infrastructure. There are several targeted attacks against ASR and AVI systems that exist in the current literature. However, none of these consider the limitations of the dissident (near-real-time, no access/knowledge of the state's  ASR and AVI systems, success over the telephony network, limited queries, transferable, high audio quality). Targeted attacks either require white-box knowledge~\cite{cisse2017houdini, yuan2018commandersong,carliniL2,schonherr2018adversarial,kreuk2018fooling} generate noisy audio~\cite{biometrics, carlini2016hidden}, are query intensive~\cite{alzantot2018did, taori2018targeted}, or not resistant to the churn of the telephone network. For a comprehensive overview of the state of the current attacks with respect to our own, we refer the reader to Table~\ref{tab:current_attacks_table} in the Appendix.

In this work, we propose the first near-real-time, black-box, model agnostic method to help evade the ASR and AVI systems employed as part of the mass telephony surveillance infrastructure\footnote{Recently, we have seen a number of attack papers against ASR and AVI systems. To better understand why our work in novel and clearly differentiate it from existing literature, we encourage the readers to review Table~\ref{tab:current_attacks_table} in the Appendix.}. Using our method, a dissident can force any ASR system to mistranscribe their phone call audio and an AVI system to misidentify their voice. As a result, governments will lose trust in their surveillance models and invest greater resources to account for our attack. Additionally, by forcing mistranscriptions, our attack will prevent governments from successfully flagging the conversation. Our attack is \emph{untargeted} i.e., it can not generate selected words or specific speakers. However, in the absence of any technique that can attain these goals in the severely constrained setting of the dissident, our method’s ability to achieve a limited set of goals (i.e., evasion) is still valuable. This can be used by dissident as the first line of defense. 

Our attack is specifically designed to address the needs of the dissident attempting to evade the audio surveillance infrastructure. The following are the key contributions of our work:

\begin{itemize}
  
 \item {\bf Our attack can circumvent any state of the art ASR and AVI system in near
	 	real-time, black-box, transferable manner:} A dissident attempting to evade mass surveillance will not have access to the target ASR or AVI systems. A key
	 	contribution of this work is the ability to generate audio
	 	samples that will induce errors in a variety ASR and AVI
	 	systems in a \emph{black-box} setting, where the adversary has no knowledge of the target model. Current black-box attacks against audio models~\cite{alzantot2018did} use genetic algorithms, which still require hundreds of thousands of queries and days of execution to generate a single attack sample. In contrast, our attack can generate a sample in \emph{fewer than 15 queries} to the target model. Additionally, we show that if dissident can not query the target model, which is most likely the case, our adversarial audio samples will still be transferable i.e., evade \emph{unknown} models.


\item {\bf Attack does not significantly impact human-perceived
        quality or comprehension and works real audio environments: } The dissident must be confident that the attack perturbations will maintain the quality of the phone call audio, survive the churn of the telephony network and still be able to evade the ASR and AVI systems. Therefore, we design our attack to introduce imperceptible changes, such that there is no significant degradation of the audio quality. We substantiate this claim by conducting an Amazon Turk user study. Similarly, we test our attacks over the cellular network, which introduces significant audio quality degradation due to transcoding, jitter and packet loss. We show that even after undergoing such serious degradation and loss, our attack audio sample is still effective in tricking the target ASR and AVI systems. To our knowledge, our work is the first to generate attack audio that is robust to the cellular network. Therefore, our attack ensures that the dissenter's adversarial audio will not have any significant impact on quality and will evade the surveillance models after having being intercepted within telephony networks.



\item {\bf Robust to existing adversarial detection and defense mechanisms:} Finally, 		   we evaluate our attack against existing techniques detecting or defending
		adversarial samples. For the former, we test the attack against the temporal based method, which has shown excellent results against traditional adversarial attacks~\cite{yang2018characterizing}. We show that this method has limited effectiveness for our attack. It is not better than randomly choosing whether an attack is in progress or not. Regarding defenses, we test our attack against adversarial training, which has shown promise in the adversarial image space~\cite{madry2017towards}. We observe that this method slightly improves model robustness, but at a cost of significant decrease in model accuracy.

\end{itemize}

The remainder of this paper is organized as follows:
Section~\ref{sec:background} provides background information on topics
ranging from signal processing to phonemes;
Section~\ref{sec:methodology} details our methodology, including our
assumptions and hypothesis; Section~\ref{sec:setup} presents our
experimental setup and parameterization; Section~\ref{sec:results} shows
our results; Section~\ref{sec:discussion} offers further discussion;
Section~\ref{sec:related} discusses related work; and
Section~\ref{sec:conclusion} provides concluding remarks.


\section{Background}
\label{sec:background}
\begin{figure}
  \includegraphics[width=\columnwidth]{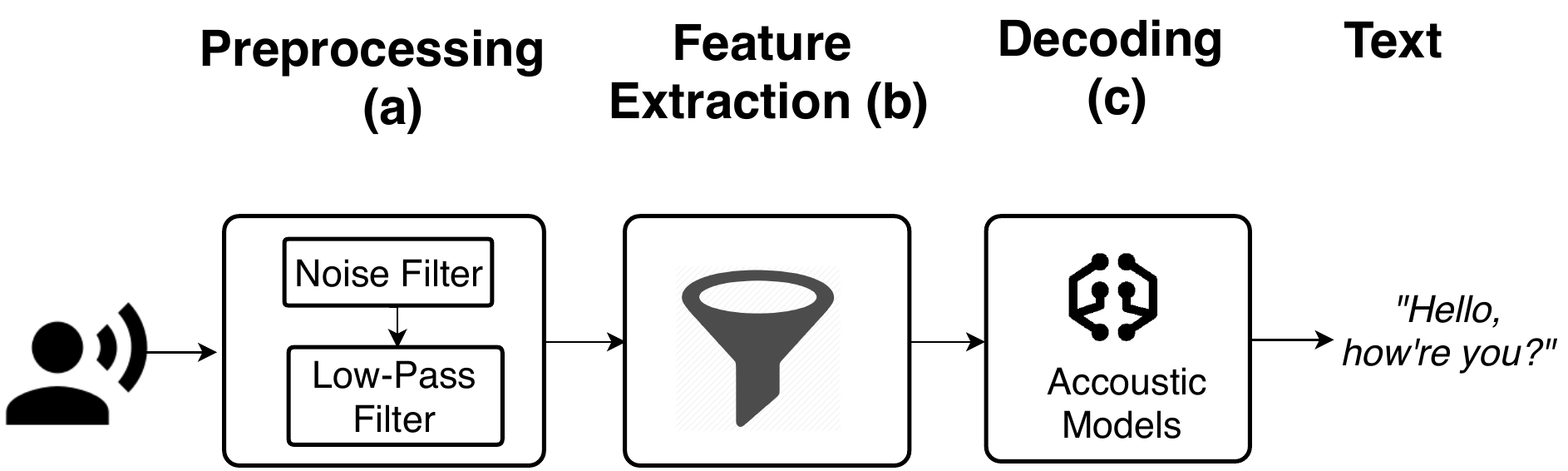}
  \vspace{-1em}
  \caption{Modern ASR systems take several steps to convert
      speech into text. (a) Preprocessing removes high frequencies and
      noise from the audio waveform, (b) feature extraction extracts the most
  important features of the audio sample, and (c) decoding converts the features
  into text.} \vspace{-2em}
  \label{fig:asr_breakdown}
  \vspace{.4em}
  \hrulefill
  \vspace{-1.8em}
\end{figure}

\subsection{Automatic Speech Recognition (ASR) Systems:}
\label{subsec:ASR}
An ASR system converts a sample of speech into text using the steps seen in
Figure~\ref{fig:asr_breakdown}.

\smallskip
\noindent
{\textbf{(a) Preprocessing}} Preprocessing in ASR systems attempts
to remove noise and interference, yielding a ``clean'' signal.
Generally, this consists of noise filters and low pass filters. 
Noise filters remove unwanted frequency components from the signal that
are not directly related to the speech.  The exact process by which the
noise is identified and removed varies among different ASR systems.
Additionally, since the bulk of frequencies in human speech fall between
300 Hz and 3000Hz~\cite{masking_book}, discarding higher frequencies
with a low pass filter helps remove unnecessary information from the
audio.



\noindent
{\textbf{(b) Feature Extraction}}\label{para:feature_extraction}
Next, the signal is converted into overlapping segments, or frames, each of
which is then passed through a feature extraction algorithm. This algorithm
retains only the salient information about the frame. A variety of signal
processing techniques are used for feature extraction, including Discrete
Fourier Transforms (DFT), Mel Frequency Cepstral Coefficients (MFCC), Linear
Predictive Coding, and the Perceptual Linear Prediction
method~\cite{rabiner1978digital}. The most common of these is the MFCC method,
which is comprised of several steps. First, a magnitude DFT of an audio sample is
taken to extract the frequency information. Next, the Mel filter is applied to
the magnitude DFT, as it is designed to mimic the human ear. This is followed by
taking the logarithm of powers, as the human ear perceives loudness on a
logarithmic scale.  Lastly, this output is given to the Discrete Cosine
Transform (DCT) function that returns the MFCC coefficients.

Some modern ASR systems use data-driven learning techniques to establish which
features to extract. Specifically, a machine learning layer (or a completely new
model) is trained to learn which features to extract from an audio sample in
order to properly transcribe the speech~\cite{venugopalan2014translating}.

\noindent
{\textbf{(c) Decoding}}
During the last phase, the extracted features are passed to a decoding function,
often implemented in the form of a machine learning model. This model has been
trained to correctly decode a sequence of extracted features into a sequence of
characters, phonemes, or words, to form the output transcription. 
ASR systems employ a variety of statistical
models for the decoding step, including Convolutional Neural Networks
(CNNs)~\cite{sainath15,ICASSP12,sainath13}, Recurrent Neural Networks
(RNNs)~\cite{graves13,sak14,sak_vinalys14,sak15}, Hidden Markov Models
(HMMs)~\cite{graves2014towards,sak15}, and Gaussian Mixture Models
(GMMs)~\cite{sphinx,sak15}. Each model type provides a unique set of properties
and therefore the type of model selected directly affects the ASR system quality.
Depending on the model, the extracted features may be re-encoded
into a different, learnable feature space before proceeding to the decoding
stage. A recent innovation is the paradigm known as \emph{end-to-end} learning,
which combines the entire feature extraction and decoding phase into one model,
and greatly simplifies the training process. The most sophisticated methods will
leverage many machine learning models during the decoding process. For example,
one may employ a dedicated language model, in addition to the decoding function,
to improve the ASR's performance on high-level grammar and rhetorical
concepts~\cite{pmlr-DS2}. Our attacks are agnostic to how the target ASR system
is implemented, making our attack completely black-box. To our knowledge, we are
the first paper to introduce black-box attacks in a limited query environment.

\subsection{Automatic Voice Identification (AVI) Systems:}
\label{subsec:spkr_rec}

AVI systems are trained to recognize the speaker of a voice
sample. The modern AVI pipeline is mostly similar to the one
used in the ASR systems, shown in Figure~\ref{fig:asr_breakdown}. While both
systems use the preprocessing and feature extraction steps, the difference lies in
the decoding step. Even though the underlying statistical model (i.e., CNNs,
RNNs, HMMs or GMMs) at the decoding stage remains the same for both systems,
what each model outputs is different. In the case of ASR systems,
the decoding step converts the extracted features into a sequence of characters,
phonemes, or words, to form the output transcription. In contrast, the decoding
step for AVI models outputs a single label which corresponds to
an individual. AVI systems are commonly used in security critical domains as an authentication mechanism to verify
the identity of a user. In our paper, we present attacks to
prevent the AVI models from correctly identifying speakers.
To our knowledge, we are the first to do so in a limited query, black-box
setting.

\subsection{Data-Transforms} In this paper, we use standard signal processing
transformations to change the representation of audio samples. The transforms can
be classified into two categories: data-independent and data-dependent. 

\subsubsection{Data-Independent Transforms}
\begin{figure}
  \centering
  \includegraphics[width=0.4\textwidth]{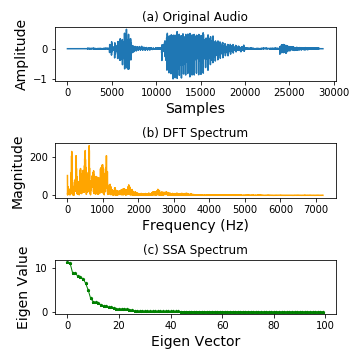}
  \caption{{\bf (a)} Original audio 
  ``about''~\cite{1k_words}; {\bf (b)} the corresponding DFT and {\bf
    (c)} SSA decompositions.  In both, low magnitude components
  (frequencies or eigenvectors, respectively) contribute little to the
  original audio.
} 
  \label{fig:1audio_2output}
    \vspace{-1.5em}
    \hrulefill
    \vspace{-1.6em}
\end{figure}

These represent input signals in terms of fixed
basis functions (e.g., complex exponential functions, cosine functions, wavelets).
Different basis functions extract different information about the signal being
transformed. For our attack, we focus on the DFT,
which exposes frequency information. We do so because the DFT is well understood
and commonly used in speech processing, both as a stand-alone tool, and as part
of the MFCC method, as discussed in Section~\ref{para:feature_extraction}.

The DFT, shown in
Figure~\ref{fig:1audio_2output}(b), represents a discrete-time series
$x_0,x_1,\ldots,x_{N-1}$ via its frequency spectrum ---~a sequence of
complex values $f_0,f_1,\ldots,f_{N-1}$ that are computed as
\(
f_k = \sum_{n=0}^{N-1} x_n \exp\left((-j2\pi)\frac{k}{N} n\right)
\)
where $j=\sqrt{-1}$, for $k=0,1,\ldots,N-1$.  One can view~$f_k$ as
the projection of the time series onto the~$k$-th basis function, a
(discrete-time) complex sinusoid with frequency $k/N$ (i.e., a sinusoid
that completes~$k$ cycles over a sequence of~$N$ evenly spaced samples). 
\ignore{
The transform is invertible, so that time series can be
written as (loosely) the sum of its projections.\footnote{For ease of
  exposition, we dispense with details concerning assumptions on the
  value of $x_n$ when $n \in \mathbb{Z}\setminus\{0,1,\ldots,N-1\}$.}  More formally, the
$n$-th time-series sample~$x_n$ is the weighted, linear combination
\(
x_n = \frac{1}{N}\sum_{k=0}^{N-1} f_k \exp\left( (j2\pi) \frac{k}{N} n\right)
\)
for $n=0,1,\ldots,N-1$
so that the~$f_k$ are sometimes referred to as the (discrete) Fourier
coefficients of the time series.
}
Intuitively, the complex-valued $f_k$ describes ``how much'' of the time
series~$x_0,x_1,\ldots,x_{N-1}$ is due to a sinusoidal waveform with
frequency $k/N$.  It compactly encodes both the magnitude of the
$k$-th sinusoid's contribution, as well as phase information, which
determines how the $k$-th sinusoid needs to be shifted (in time).
The DFT is invertible, meaning that
a time-domain signal is uniquely determined by a given sequence of
coefficients.  Filtering operations (e.g. low-pass/high-pass filters)
allow one to accentuate or downplay the contribution of specific
frequency components; in the extreme, setting a non-zero $f_k$ to zero
ensures that the resulting time-domain signal will not contain that
frequency.  

\ignore{
The DFT converts an $N$-sample time-domain signal
into an $N$-vector of complex-valued coefficients~$F_k$, each compactly
encoding the amplitude and phase of an associated complex sinuoid. corresponding to a sinusoidal signal
with a fixed frequency. Figure~\ref{fig:1audio_2output}(b) shows the DFT
transformation of an audio sample shown in Figure~\ref{fig:1audio_2output}(a),
using the following equation:
\begin{gather*}
    F_k = \sum_{n=1}^{N-2}s_n [\mathrm{cos}[\frac{\pi}{N}(n+\frac{1}{2})k] 
    -i \cdot \mathrm{sin}[\frac{\pi}{N}(n+\frac{1}{2})k]] \\
    \quad \quad k=0,...,N-1
  \nonumber 
\end{gather*}
where $F = (F_1,...,F_N)$ is a vector of
complex numbers, $N$ is the total number of samples, $n$ is the sample number,
$k$ is the frequency number, and $s_n$ is the n-th sample of $S$.  The magnitude of
$F_k$ corresponds to the intensity of the $k^{th}$ frequency
in $s$. Studying the DFT output provides a fine granularity insight into an
audio sample's spectral makeup. The DFT is invertible, meaning that
the sequence $f_k$; this allows filtering of
frequency components from the original signal. Namely, by setting $f_k = 0$
prior to DFT inversion, the time-domain signal will not contain any contribution
at that frequency.
}

\subsubsection{Data-Dependent Transforms} Unlike the
DFT, data-dependent transforms do not use predefined basis functions. Instead,
the input signal itself determines the effective basis functions: a set
of linearly independent vectors which can be used to
reconstruct the original input. Abstractly, an input sequence $\mathbf{x}$
with $|\mathbf{x}| = n$ can be thought of as a vector in the space
$\mathbb{R}^n$, and the data-driven transform finds the bases for the input
$\mathbf{x}$. 
Singular Spectrum Analysis (SSA) is a spectral estimation method that decomposes
an arbitrary time series into components called \emph{eigenvectors}, shown in
Figure~\ref{fig:1audio_2output}(c). These eigenvectors represent the various
trends and noise that make up the original series. Intuitively, eigenvectors
corresponding to eigenvalues with smaller magnitudes convey relatively less
information about the signal, while those with larger eigenvalues capture
important structural information, as long-term ``shape'' trends, and dominant
signal variation from these long-term trends. Similar to the DFT, the SSA is also linear and invertible.
Inverting an SSA decomposition after discarding eigenvectors with small
eigenvalues is a means to remove noise from the original series.

\subsection{Cosine Similarity} Cosine Similarity is a metric used to measure the
similarity of two vectors. This metric is often used to measure how similar two
samples of text are to one another (e.g., as part of the TF-IDF
measure~\cite{kopcke2010evaluation}). In order to calculate this, the sample
texts are converted into a dictionary of vectors. Each index of the vector
corresponds to a unique word, and the index value is the number of times the
word occurs in the text. The cosine similarity is calculated using the equation 
\(cos( x,  y) = \frac { x \cdot  y}{|| x|| \cdot || y||}\),
\noindent where $x$ and $y$ are the sentence samples. Cosine values close to one mean that
the two vectors, or in this case sentences, have high similarity.

\subsection{Phonemes}

Human speech is made up of various component sounds known as phonemes. The set
of possible phonemes is fixed due to the anatomy that is used
to create them. The number of phonemes that make up a given language
varies. 
English, for example, is made up of 44 phonemes. Phonemes can be divided into
several categories depending on how the sound is created. These categories
include vowels, fricatives, stops, affricates, nasal, and glides. In this paper,
we mostly deal with fricatives and vowels; however, for completeness, will
briefly discuss the other categories here.

\textit{Vowels} are created by positioning the tongue and jaw such that two resonance
chambers are created within the vocal tract. These resonance chambers create
certain frequencies, known as formants, with much greater amplitudes than
others. The relationship between the formants determines which vowel is heard.
Examples of vowels include $iy$, $ey$, and $oy$ in the words beet, bait, and
boy, respectively.

\textit{Fricatives} are generated by creating a constriction in the airway that causes
turbulent flow to occur. Acoustically, fricatives create a wide range of higher
frequencies, generally above 1 kHz, that are all similar in intensity.  Common
fricatives include the $s$ and $th$ sounds found in words like sea and thin.

\textit{Stops} are created by briefly halting air flow in the vocal tract before
releasing it. Common stops in English include $b$, $g$, and $t$ found in
the words bee, gap, and tea. Stops generally create a short section of silence
in the waveform before a rapid increase in amplitude over a wide range of
frequencies.

\textit{Affricates} are created by concatenating a stop with a fricative. This results in
a spectral signature that is similar to a fricative.  English only contains two
affricates, $jh$ and $ch$ which can be heard in the words joke and chase,
respectively.

\textit{Nasal} phonemes are created by forcing air through the nasal cavity. Generally,
nasals have less amplitude than other phonemes and consist predominantly
of lower frequencies. English nasals include $n$ and $m$ such as in the words
noon and mom.

\textit{Glides} are unlike other phonemes, since they are not grouped by their means of
production, but instead by their roll in speech. Glides are acoustically similar
to vowels but are instead used like consonants, acting as transitions between
different phonemes. Examples of glides include the $l$ and $y$ sounds in lay and
yes.



 

\section{Methodology}\label{sec:methodology}
\subsection{Hypothesis and Threat Model}\label{sub:threat_model}

\textit{Hypothesis:}
Our central hypothesis is that ASR and AVI systems rely on components of speech
that are non-essential for human comprehension. Removal of these components can
dramatically reduce the accuracy of ASR system transcriptions and AVI system 
identifications without significant loss of audio clarity. Our methods and experiments 
are designed to test this hypothesis.

\textit{Threat Model and Assumptions:}
For the purposes of this paper, we define the attacker or adversary as a person who is aiming
to trick an ASR or AVI system via audio perturbations. In contrast, we define the defender 
as the owner of the target system.

We assume the attacker has no knowledge of the type, weights, architecture, or
layers of the target ASR or AVI system. Thus, we treat each system as a black-box to which we make
requests and receive responses. We also assume the attacker can only make a
limited number of queries to the target model, as a large number of queries will
alert the defender of the attacker's activities. Furthermore, the
attacker has less computational resources than the defender.



We assume the defender has the ability to train an ASR or AVI system. Additionally,
they may use any type of machine learning model, feature extraction, or
preprocessing to create their ASR or AVI system. Finally, the defender is able to monitor incoming queries to their system and prevent attackers from performing large numbers of queries.

\subsection{Attack Steps}\label{sub:attack_steps}
\begin{figure}
  \includegraphics[width=0.48\textwidth]{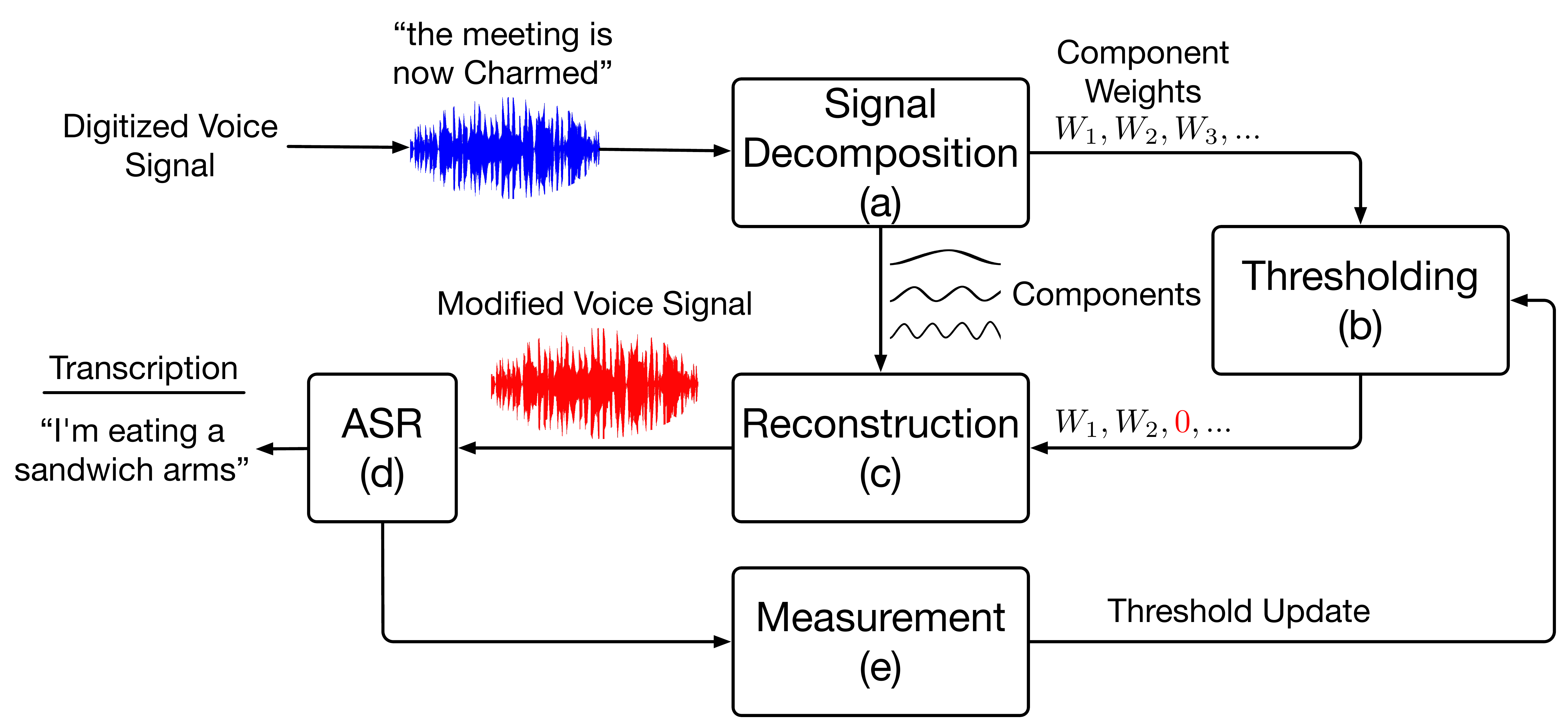}
  \caption{The figure shows the steps involved in generating an attack audio
  sample. First, the target audio sample is passed through a signal decomposition
  function (a) which breaks the input signal into components. Next, subject to
  some constraints, a subset of the components are discarded during thresholding
  (b). A perturbed audio sample is reconstructed (c) using the remaining weights
  from (a) and (b) . The audio sample is then passed to the ASR/AVI system (d) for
  transcription. The difference between the transcription of the perturbed audio
  and the original audio is measured (e). The thresholding constraints are
  updated accordingly (c) and the entire process is repeated.  }
  \label{fig:single_attack_step}
 \vspace{-1.5em}
 \hrulefill
 \vspace{-1.6em}
\end{figure}

Readers might incorrectly assume that certain trivial attacks might be able to
achieve the goals of the dissident i.e., evade the model while maintaining high 
audio quality. One such trivial attack includes adding white-noise to the speech
samples, expecting a mistranscription by the model. However, such a 
an attack will fail. We discuss in detail how we test this trivial attack and the 
corresponding results in Appendix~\ref{sub:white_attack}.
Similarly we introduce a simple impulse
perturbation technique that exposes the sensitivity of ASR systems, discussed in Appendix~\ref{sub:impulse_attack_meth}.  Realizing
the limitations of this approach, we leave its details to
Appendix~\ref{sub:impulse_attack_meth} to~\ref{para:impulse_attack_discussion}.
We continue our study to develop a more robust attack algorithm in the following
sections. 


The attack should meet certain constraints: First, it should introduce
artificial noise to the audio sample that exploits the ASR/AVI system and
forces an incorrect model output.
Second, the distortion should have little to
no impact on the understandability of the perturbed audio file for the human
listener.

The attack steps are outlined in Figure~\ref{fig:single_attack_step}. During
decomposition, shown in Figure~\ref{fig:single_attack_step}(a),
we pass the audio sample to the selected algorithm (DFT or SSA).  The algorithm
decomposes the audio into individual components and their corresponding
intensities. Next, we threshold these components, as shown in
Figure~\ref{fig:single_attack_step}(b).
During thresholding, we remove the components whose intensity falls below a
certain threshold. The intuition for doing so is that the lower intensity
components are less perceptible to the human ear and will likely not affect a
user's ability to understand the audio. We discuss how the algorithm calculates
the correct threshold in the next paragraph. We then construct a new audio
sample from the remaining components, using the appropriate inverse transform,
shown in Figure~\ref{fig:single_attack_step}(c). Next, the audio is passed on
to the model for inference, Figure~\ref{fig:single_attack_step}(d). If the
system being attacked is an ASR, then the model outputs a transcription. On the
other hand if the target system is an AVI, the model outputs a speaker label.
The model output is
compared with that of the original during the measurement step,
Figure~\ref{fig:single_attack_step}(e). 

The goal of the algorithm is to calculate the optimum threshold,
which discards the least number of components whilst still forcing the model to
misinterpret the audio sample. Discarding components degrades the quality of
the reconstructed audio sample. If the discard threshold is too high,
neither the human listener nor the model will be able to correctly interpret the
reconstructed audio. On the other hand, by setting it too low, both the human
listener and the model will correctly understand the reconstructed audio.

To compensate for these competing tensions, the attack algorithm executes the
following steps. If the model output matches the original label, during the measurement step, the
algorithm will increase the threshold value. It will then pass the audio sample
and the updated threshold back to the thresholding step for an additional round
of optimization. However, if the model outputs an incorrect interpretation of
the audio sample, the algorithm reduces the degradation by reducing the discard
threshold, before returning the audio to the thresholding step. This loop will
repeat until the algorithm has calculated the optimum discard threshold.
\vspace{-1em}
\subsection{Performance}\label{sub:improved_attack_steps}
In order to find the optimal threshold, we incrementally remove
more components until the model fails to properly transcribe the audio file.
This process takes $O(n)$ queries to the model, where $n$ is the number of
decomposition components. We can reduce the time complexity from linear to
logarithmic time such that an attack audio is produced in $O(\log~n)$ queries. To
achieve this, we model the distortion search as a binary search 
(BS) problem where values represent the number of coefficients to use
during reconstruction.

If the reconstruction is misclassified, we move to the left BS search-space and
attempt to improve the audio quality by removing less coefficients. If the
audio is correctly transcribed, we move to the right. This search continues
until either an upper bound on the search depth is reached. This result was
sufficient for the scope of this paper, and we leave a more rigorous analysis
of distortion search complexity for future work.

\subsection{Transferability}\label{sub:improved_attack_steps}
One measure of an attack's strength is the ability to generate adversarial
examples that are transferable to other models (i.e., a single audio sample that is
mistranscribed/misidentified by multiple models). An attacker will not
know the precise model he is trying to fool. In such a scenario, the attacker will
generate examples for a known model, and hope that the samples will work against
a completely different model.

Attacks have been shown to generalize between models in the image
domain~\cite{papernot17}. In contrast, attack audio transferability has only
has seen limited success. Additionally, audio generated with previous
approaches (\cite{carlini2016hidden,carliniL2}) are sensitive to naturally
occurring external noise, which fails to exploit the target model in a
real-world setting. This is in line with previous results of physical attacks
in the image domain~\cite{2017arXiv170703501L}. Instead we focus on the
\emph{evasion} style of attack, where the attack is considered successful if
the ASR system transcribes the attack audio incorrectly or the AVI
misidentifies the speaker of the attack audio. We propose a completely
black-box approach that does not consider model specifics as a means of
bypassing these limitations. 

\subsection{Detection and Defense}\label{sub:detection_and_defense}
We evaluate our attack against the adversarial audio detection mechanism which is based on temporal dependencies~\cite{yang2018characterizing}. This is the only method designed specifically to detect adversarial audio samples. This method has demonstrated excellent results: it is light-weight, simple and highly effective at detecting tradition adversarial attacks. The mechanism takes as input an audio sample. This can either be adversarial or benign. Next, the audio sample is partitioned into two. Only the partition corresponding to the first half is retained. Next, the entire original audio sample and the first partition are passed to the model and the transcriptions are recorded. If the transcriptions are similar, then the audio sample is considered benign. However, if the transcriptions are differing, the audio sample is adversarial. This is because adversarial attack algorithms against audio samples distort the temporal dependence within the original sequences. The temporal dependency-based detection is designed to capture this information and use it for attack audio detection.

Additionally, we evaluate our attack against adversarial training based defense. However, we have placed the steps, the methodology, and the results in the Appendix~\ref{sub:adv_training}.


\section{Setup}\label{sec:setup}
Our experimental setup involved two datasets, four attack scenarios,
two test environments, seven target models, and a user study. We discuss the
relevant details of our experiments here.

\subsection{TIMIT Data Set}\label{sub:timit}
The TIMIT corpus is a standard dataset for speech processing systems. It
consists of 10 English sentences that are phonetically diverse being spoken by
630 male and female speakers~\cite{garofolo1988getting}. Additionally, there is
metadata of each speaker that includes the speaker's dialect. In our tests, we
randomly sampled six speakers, three male and three female, from each of four
regions (New England, Northern, North Midland, and South Midland). We then
perturbed all 10 sentences spoken by our speakers using our technique, and also
extracted phonemes for our phoneme-level experiments. In total, we attacked 240
recordings with 7600 phonemes.

\subsection{Word Audio Data Set}\label{sub:word_audio_data_set}


Testing the word-level robustness of an ML model poses challenges in terms of
experimental design. Although there exist well-researched datasets of spoken
phrases for speech transcription~\cite{panayotov2015librispeech,an4_db},
partitioning the phrases into individual words based on noise threshold is not
ideal. In this case, the only way to control the distribution of candidate
phrases would be to pass them to a strong transcription model, while discarding
audio samples which are mistranscribed. Doing so may bias the candidate attack
samples towards clean, easy to understand samples. Instead, we build a
word-level candidate dataset using a public repository of the 1,000 most common
words in the English language~\cite{1k_words}. We then download audio for each
of the 1,000 words using the Shotooka spoken-word service~\cite{shotooka}.

\subsection{Attack Scenarios}\label{sub:attacks}
In order to test our technique in a variety of different ways, we performed four
attacks: word level, phoneme level, ASR Poisoning, and AVI
poisoning. We also tested the transferability of the attack audio samples across
models.



\subsubsection{Word Level Perturbations}\label{sub:word_level_perturbations}
Using the 1,000 most common words, we performed our attack as described in
\mbox{Section}~\ref{sub:attack_steps}.
We optimized our threshold using the technique outlined in Section
\ref{sub:improved_attack_steps}, stopping either after the threshold value had
converged or after a maximum of 15 queries.

\subsubsection{Phoneme Level Perturbations}\label{sub:phoneme_level_perturbations}
Next, we ran perturbations on individual phonemes rather than entire words. The
goal of this attack was to cause mistranscription of an entire word by only
perturbing a single phoneme. We tested this attack on audio files from the TIMIT
corpus and replaced the regular phoneme with its perturbed version in the audio
file. The audio sample was then passed to the ASR system for
transcription. We repeated this process for every phoneme using the binary search
technique outlined in \mbox{Section}~\ref{sub:improved_attack_steps}.

\subsubsection{ASR Poisoning}\label{sub:sentence_poinsoning}
ASR systems often use the previously transcribed words to infer the
transcription of the
next~\cite{chorowski2015attention,rabiner1993fundamentals,bahdanau2016end}.
Perturbing a single word's phonemes not only affects the model's ability to
transcribe that word, but also the words that come after it. We wished to
measure the effect of perturbing a single phoneme on the transcription of the
remaining words in a sentence. To do this, we generated adversarial audio sample
by perturbing a single phoneme while keeping the remaining audio intact for the
sentences of the TIMIT dataset. We repeated this for every phoneme in the
dataset and passed the attack audio samples to the ASR system. The cosine
similarity metric was used to measure the transcription similarity between the
attack audio with the original audio.


When perturbing phonemes, we do not use the attack optimization described in
Section~\ref{sub:improved_attack_steps}. Since the average length of a phoneme
in our dataset was only 31ms, a single perturbed phoneme in a sentence does not
significantly impact audio comprehension. Therefore, we simply discard half of all
decomposition coefficients in a single 31 ms window during the thresholding step. This maintains the
quality of the adversarial audio, while still forcing the model to mistranscribe.

\subsubsection{AVI Poisoning} 
\label{sub:sr_poisoning}
We also evaluate our attacks' performance against AVI system. To do so, we first
trained an Azure Speaker Identification model~\cite{azure_attest} to recognize 20 speakers. We selected 10 male and 10
female speakers from the TIMIT dataset to service as our subjects.  For each
speaker, seven sentences were used for training, while the remaining three
sentences were used for attack evaluation. We only perturbed a single
phoneme while the rest of the sentence is left unaltered. We passed both the
benign and adversarial audio samples to the model. The attack was considered a
success if the AVI model output different labels for each sample. This attack setup is similar to the one for ASR poisoning, except that here we target an AVI system. 
\subsection{Models}\label{sub:models}

We choose a set of seven models that are representative of the state-of-the-art
in ASR and AVI technology, given their high
accuracy~\cite{article_acc_models,wer_are_we,google_speech_error,danielsson2017}.
These include a mixture of both proprietary and open-source models to expose the
utility of our attack. However, all are treated as black-box.

\smallskip
\noindent
{\bf Google (Normal):} To demonstrate our attack in a truly
black-box scenario, we target the speech transcription APIs provided by Google.
The `Normal' model is provided by
Google for `clean' use cases, such as in home assistants,
where the speech is not expected to traverse a cellular
network~\cite{google_normal}.

\smallskip
\noindent
{\bf Google (Phone):} To demonstrate our attack against model trained
for noisy audio, we test the attack against the `Phone' model.  Google provides
this model for cellular use cases and trained it on call audio that will
be representative of cellular network compression~\cite{google_phone}. We also
assume that the Google `Phone' model will be robust against the noise, jitter,
loss and compression introduced to audio samples that have traveled through the
telephony network. 

\smallskip
\noindent
{\bf Facebook Wit:} To ensure better coverage across the space of
proprietary speech transcription services, we also target Facebook
Wit, which provides access to a `clean' speech transcription model~\cite{wit}. As before,
no information is known about this model due to its proprietary nature.

\smallskip
\noindent
{\bf Deep Speech-1:} The goal of Deep Speech
1 was to eliminate hand-crafted feature pipelines by leveraging a
paradigm known as \emph{end-to-end learning}~\cite{graves2014towards,hannun2014deep}. This
results in robust speech transcription despite noisy environments, if sufficient
training data is provided. For our experiments, we use an open-source
implementation and checkpoint provided by Mozilla with MFCC
features~\cite{mozilla_ds}.

\smallskip
\noindent
{\bf Deep Speech-2:} Deep Speech-2 introduced architecture
optimizations for very large training sets. It is trained to map raw audio
spectrograms to their correct transcriptions, and demonstrates the current
state-of-the-art in noisy, end-to-end audio transcription~\cite{pmlr-DS2}. We
use an open-source implementation\footnote{At the time of running our
experiments, the implementation did not include a language model to aid in the
beam-search decoding.} trained on LibriSpeech provided by GitHub user
SeanNaren~\cite{panayotov2015librispeech,ds2_pytorch}. The primary difference in
our two tested versions is feature preprocessing: the tested version of Deep
Speech-1 uses MFCC features, while the tested version of Deep Speech-2 uses raw
audio spectrograms.

\smallskip
\noindent
{\bf CMU Sphinx:} The CMU Sphinx project is an open-source speech
transcription repository representing over twenty years of research in this
task~\cite{sphinx}. Sphinx does not heavily rely on deep learning techniques,
and instead implements a combination of statistical methods to model speech
transcription and high-level language concepts. We use the PocketSphinx
implementation and checkpoints provided by the CMU Sphinx
repository~\cite{sphinx}.

\smallskip
\noindent
{\bf Microsoft Azure:} To demonstrate our attack against AVI 
systems in a black-box environment, we attack the Speaker
Identification API provided by Microsoft Azure~\cite{azure_attest}. This system is
proprietary, and hence, completely black-box. There is no publicly available
information about the internals of the system.

Although the landscape of audio models is ever-changing, we
believe our selection represents an accurate approximation of both traditional
and state-of-the-art methods. Intuitively, future models will derive from the
existing state-of-the-art in terms of data and implementation. We also compare
against open-source and proprietary models, to show our approach generalizes to
any model regardless of a lack in \emph{apriori} knowledge.

\subsection{Transferability}\label{sub:transferability}


\newcommand{\overbar}[1]{\mkern
1.5mu\overline{\mkern-1.5mu#1\mkern-1.5mu}\mkern 1.5mu} We measure the
transferability of our proposed attack by finding the probability that the
attack samples for one model will successfully exploit another. This is done by
creating a set of successful word-level attack audio samples $X^*_m$ for a
model $m$, then averaging their calculated Mean Squared Error (MSE) distortion,
$\overbar{MSE_m}$. Intuitively, this average MSE will be higher for stronger
models, and lower for weaker models. This acts as a `hardness score' for a
given model and is used to compare between attack audio sets of two models. Now
consider a baseline model $f$, comparison model $g$, the successful attack
transfer event $S_{f \rightarrow g}$, and the number of audio samples in each
model's attack audio set $n = |X^*_f| = |X^*_g|$. We can calculate attack
transfer probability from $f$ to $g$ as the probability of sampling attack
audio from $X^*_f$ whose distortion meets or surpasses the score
$\overbar{MSE_g}$. We denote this probability $P(S_{f \rightarrow g})$ and the
set of potentially transferable audio samples as $V_{f \rightarrow g}$. We
calculate the probability using the equation
	\(P(S_{f \rightarrow g}) =  \frac{|V_{f \rightarrow g}|}{n}\),
\noindent where we build the set of transferable attack samples such that
	 \(V_{f \rightarrow g} = \{\textbf{x}^*_{f,i} \in X^*_f : MSE(\textbf{x}^*_{f,i}) \geq \overbar{{MSE}_g}\}\).

Thus, we approximate the probability of sampling a piece of audio which meets
the `hardness' of model $g$ from the set which was successful over model $f$.
This value is calculated across each combination of models in our experiments
for SSA and DFT transforms, with $n$ set to 1,000 as a result of using our Word
Audio data set.

\subsection{Detection}\label{sub:Detection_meth}
Our experimental method was designed to be as close to that of the authors~\cite{yang2018characterizing}. We assume that the attacker is not aware of the existence of the defense mechanism or the size of the partition. Therefore, we perturbed the entire audio sample using our attack, to maximally distort the temporal dependencies. Next, we partitioned the audio sample into two halves. We passed both the entire audio sample and the first half to the Google Speech API for transcription. We conducted this experiment with 266 adversarial audio samples generated using the DFT perturbation method at a factor of 0.07. Our set of benign audio samples consisted of 534 benign audio samples. The audio samples for both the benign and adversarial sets were taken from the TIMIT dataset. Similar to the authors, we use Word Error Rate (WER) as a measure of transcription similarity. In line with the authors, we calculate and report the Area Under Curve (AUC) value. AUC values lie between 0.5 and 1. A perfect detector will return an AUC of 1, while a detector that randomly guesses returns an AUC of 0.5.


\subsection{Over-Cellular}\label{sub:cellular}
The Over-Cellular environment simulates a more realistic scenario where an
adversary's attack samples have to travel through a noisy and lossy channel
before reaching the target system. Additionally, this environment accurately models
one of the most common mediums used for transporting human voice -- the
telephony network. In our case, we did this by sending the audio through AT\&T's
LTE network and the Internet via Twilio~\cite{twilio} to an iPhone 6. The
attacker's audio is likely to be distorted while passing through the cellular
network due to jitter, packet loss, and the various codecs used within the
network. The intuition for testing in this environment is to ensure that our
attacks are robust against these kinds of distortions.



\begin{figure}  
  \centering
  \includegraphics[width=1\columnwidth]{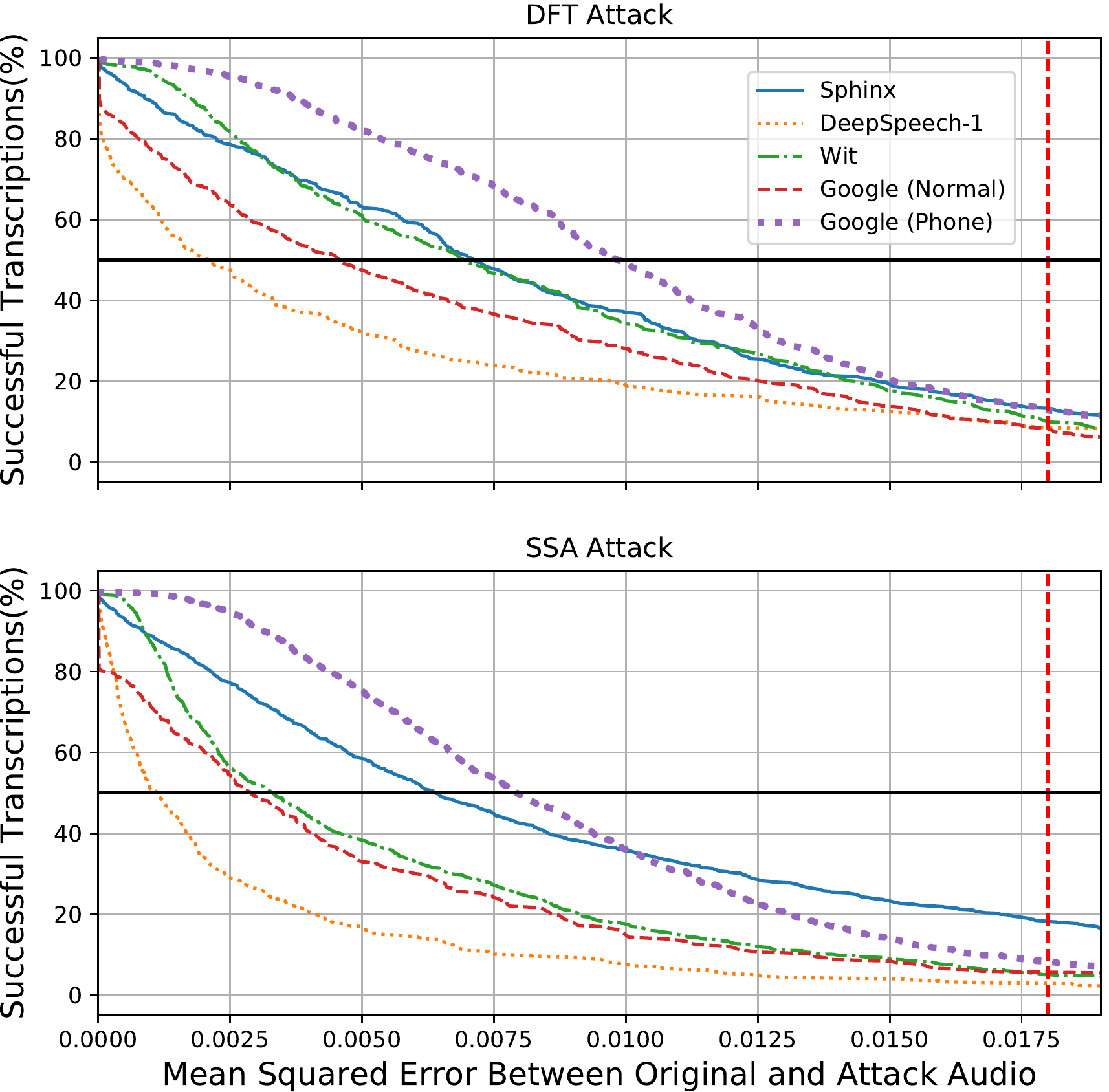}
  \caption{Success transcriptions against our word-level attack plotted against increasing
  distortion, calculated using Mean Square Error (MSE). The SSA-based
  word-level attack sees a faster, sharper decrease in the successful transcriptions than the
  DFT-based word-level attack, noted by its ability to reach 50\% attack
  success (solid black line) across all models within a smaller span of
  distortion. This means 50\% of the words in the dataset were mistrascribed by the target ASR. In every case, the test set accuracy falls considerably before
  reaching the GSM baseline distortion (dashed red line).}
  \label{fig:word_atk_MSE}
  \vspace{-1.6em}
  \hrulefill
  \vspace{-1.6em}
\end{figure}

\subsection{MTurk Study Design}\label{sub:mturk-study-design}
In order to measure comprehension of perturbed phone call audio, we conducted an
IRB approved online study.  We ran our study using Amazon's Mechanical Turk
(MTurk) crowdsourcing platform.  All participants were between the ages of $18$
and $55$ years old, located in the United States, and had an approval rating for
task completion of more than $95\%$. During our study, each participant was
asked to listen to audio samples over the phone. Parts of the audio sample had been perturbed, while others had been left unaltered\footnote{We invite the readers to listen to the perturbed audio samples
for themselves: https://sites.google.com/view/transcript-evasion}. The audio
samples were delivered via an automated phone call to each of our participants.
The participants were asked to transcribe the audio of a pre-recorded
conversation that was approximately one minute long. After the phone call was
done, participants answered several demographic questions which concluded the
study.  We paid participants \$2.00 for completing the task. Participants were
not primed about the perturbation of the pre-recorded audio, which prevented
introducing any bias during transcription.  In order to make our study
statistically sound, we ran a sample size calculation under the assumption of
following parameter values: an effect size of 0.05, type-I error rate of 0.01,
and statistical power of 0.8. Under these given values, our sample size was
calculated to be 51 and we ended up recruiting 89 participants in MTurk.
Among these 89, 23 participants started but did not complete the study and 5
participants had taken the study twice. After discarding duplicate and
incomplete data, our final sample size consists of 61 participants.

%


\section{Results}\label{sec:results}
As outlined previously in Section~\ref{sub:attacks}, we evaluated our attack in
various different configurations in order to highlight certain properties of the
attack. To begin, we will evaluate our attack against the speech to text
capabilities of multiple ASR systems in several different setups.


\subsection{Attacks Against ASR systems}
\label{sub:asr_attacks}

\subsubsection{Word Level Perturbations}\label{sub:word_level_perturbations_result}


\begin{figure*} 
\centering
  \includegraphics[width=\textwidth]{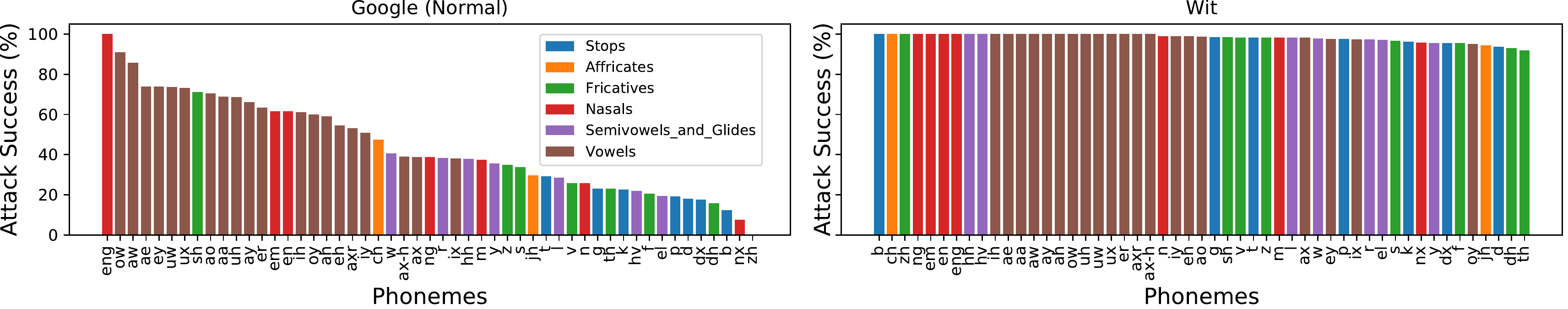}

  \caption{A comparison of attack success of our DFT-based phoneme-level attack
      against two ASR models. There is a clear relationship between the which
      phoneme is attacked and the attack's success. It is clear across all
      models that we evaluated that vowels are more vulnerable to this attack
  than other phonemes.}
  \label{fig:phoneme_atk_succ}
  \vspace{-1.6em}
  \hrulefill
\end{figure*}


We study the effect of our word-level attack against each model. We measure
attack success against distortion and compare the DFT and the SSA attacks, shown
in \mbox{Figure}~\ref{fig:word_atk_MSE}. In this subsection, we discuss five
target models: Google (Normal), Google (Phone), Wit, DeepSpeech-1 and Sphinx.
Distortion is calculated using the MSE between every normal audio
sample and its adversarial counterpart.

We use the GSM audio codec's average MSE as a baseline for audio
comprehension, as it is used by 2G cellular networks (the most common globally).
We denote this baseline with the red, vertical dashed line in
Figure~\ref{fig:word_atk_MSE}. Thus, we consider any audio with higher MSE than
the baseline to be totally incomprehensible to humans. It is important to note
that this assumption is extremely conservative, since normal comprehensible
phone call audio often has larger MSE than our baseline.

Figure~\ref{fig:word_atk_MSE} shows that as distortion is iteratively
increased using the word-level attack, test set accuracy begins to diminish
across all models and all transforms. Models which decrease slower, such as
Google (Phone), indicate a higher robustness to our attack. In contrast, weaker models, such as
Deep Speech-1, exhibit a sharper decline. For all transforms, the Google
(Phone) model outperforms the Google (Normal) model. This indicates that training the
Google (Phone) model on noisy audio data exhibits a rudimentary form of adversarial
training. However, all attacks are eventually successful to at least 85\%
while retaining audio quality that is comprehensible to humans.

Despite implementing more traditional machine learning techniques, Sphinx
exhibits more robustness than Deep Speech-1 across both attacks. This
indicates that Deep Speech-1 may be overfitting across certain words or
phrases, and its existing architecture is not appropriate for publicly
available training data. Due to the black-box nature of Wit and the Google models,
it is difficult to compare them directly to their white-box counterparts.
Overall, Sphinx is able to match Wit's performance, which is also
more robust than the Google (Normal) model in the DFT attack.

Surprisingly, for the SSA attack Sphinx is able to outperform all models as
distortion approaches the human perceptibility baseline. This may be a
byproduct of the handcrafted features and models built into Sphinx. Overall,
the SSA-based attack manages to induce less distortion, allowing all models to
fail with 50\% (represented by the horizontal black line) or less test set
accuracy before 0.0100 calculated MSE. Manual listening tested showed that there
was no perceivable drop in audio quality at this MSE.

\subsubsection{Phoneme Perturbations}\label{sub:phoneme_perturbations}

%
%

Our evaluation of the phoneme level attacks exposed several trends, as shown
in Figures~\ref{fig:phoneme_atk_succ} and ~\ref{fig:phoneme_comp}. For brevity,
only Google (Normal) and Wit are shown across each data transform, with
complete charts available in Section~\ref{app:results} of the Appendix.

Figure~\ref{fig:phoneme_atk_succ} shows the relationship between phonemes and
attack success. Lower bars correspond to a greater percentage of attack
examples that were incorrectly transcribed by the model. According to the
figure, vowels are more sensitive to perturbations than other phonemes. This
pattern was consistently present across all the models we evaluated.
There are a few possible explanations for this behavior. Vowels are the most
common phonemes in the English language and their formant
frequencies are highly structured. It is possible these two aspects in tandem
force the ASR system to over-fit to the specific formant pattern during
learning. This would explain why vowel perturbations cause the model to mistranscribe.

Similarly, Figure~\ref{fig:phoneme_comp} shows the distortion thresholds
needed for each phoneme to cause mistranscription of a phrase. The longer the
bar, the greater the required threshold.
For the DFT-based attack experiment, shown in Figure~\ref{fig:phoneme_comp}(a),
we observe that the vowels require a lower threshold compared to the other
phonemes. This means that less distortion is required for a vowel to trick the ASR
system. In contrast, the SSA-based attack experiments, as shown in
Figure~\ref{fig:phoneme_comp}(b), reveal that all of the phonemes are equally
vulnerable. In general, the SSA attacks required a higher threshold than our
previous DFT attack. However, the MSE of the audio file after being perturbed when
compared to the original audio is still small. The average MSE during these
tests was $0.0067$, which is an order of magnitude less than the MSE of audio
being sent over LTE ($0.0181$).

Our SSA attacks did not appear to expose any systemic vulnerability in our
models as the DFTs did. There exist two likely causes for this: DFT's use
in ASR feature extraction and SSA's data dependence. ASR systems often use DFTs as
part of their feature extraction mechanisms, and thus the models are likely
learning some characteristics that are dependent on the DFT. When our attack
alters the DFT, we are directly altering acoustic characteristics that will
affect which features the model is extracting and learning.

Additionally, when SSA is used in our attack, we are removing eigenvectors
rather than specific frequencies like we do with a DFT. These eigenvectors
are made up of multiple frequencies that are not unique to any one
eigenvector. Thus, the removal of an eigenvector does not guarantee the
complete removal of a given frequency from the original audio sample. We
believe the combination of these two factors results in our SSA-based attack
being equally effective against all phonemes.

Both Figures~\ref{fig:phoneme_atk_succ} and~\ref{fig:phoneme_comp} provide
information for an attacker to maximize their attack success probability.
Perturbing vowels at a 0.5 threshold while using a DFT-based attack will provide
the highest probability for success because vowels are vulnerable across all
models. Even though the discard threshold applied to vowels might vary from one
model to another, choosing a threshold value of 0.5 can guarantee both stealth
and a high likelihood of a successful mistranscription.

%



\begin{figure} 
  \includegraphics[width=0.45\textwidth]{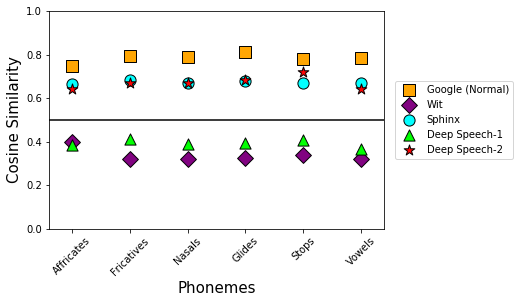}
  \caption{Cosine similarity between the transcriptions of
  the original and the perturbed audio file. At a value of 0.5 (horizontal
  line) half of the sentence is incorrect. 
  Attack audio samples were generated by
  perturbing a single phoneme. 
} 
  \label{fig:domino_global}
  \vspace{-1.5em}
  \hrulefill
  \vspace{-1.6em}
\end{figure}

\begin{table*}[t]
\centering
\vspace{-1em}
\begin{tabular}{|l|l|l|}
\hline
\textbf{Model}  & \textbf{Original Transcription}  & \textbf{Attack Transcription}      \\ \hline

\multirow{2}{*}{Google (Normal)} & \textbf{\underline{T}}he emperor had a mean Temper  & syempre Hanuman Temple \\ \cline{2-3}

  & then the ch\textbf{\underline{O}}reographer must arbitrate                & Democrat ographer must arbitrate   \\ \hline

\multirow{2}{*}{Wit}             & she had your dark suit in \textbf{\underline{G}}reasy wash water all year & nope                               \\ \cline{2-3}

                                & masquerade parties tax one'\textbf{\underline{S}} imagination             & stop                               \\ \hline

\end{tabular}
\smallskip
\smallskip

By only perturbing a
single phoneme (bold faced and underlined), our attack forces ASR systems to
completely mistranscribe the resulting audio.

\label{tbl:poison_examples}
\vspace{-1.6em}
\hrulefill
\end{table*}

\subsubsection{ASR Poisoning}\label{sub:sentence_poinsoning_result}
As described in Section~\ref{sub:phoneme_level_perturbations}, perturbing a
single phoneme not only causes a mistranscription of the given word but also of
the following words as well. Results of this phenomena can be seen in Table~\ref{tbl:poison_examples}. We
further, characterize this numerically across each model for the DFT-based attack
in Figure~\ref{fig:domino_global}, where higher values of cosine similarity
translate to lower attack mistranscription.

We observe a relationship between the model type and the cosine similarity
score. Of all the models tested, Wit is the most vulnerable, given low average
cosine similarity of 0.36. On the other hand, the Google (Normal) model seems to
be least vulnerable with the highest cosine similarity of 0.78. To better
characterize the phenomenon, we use the cosine similarity value to estimate the
number of words that the attack effects. We do so by assuming a sentence
comprised of 10 words each. Perturbing a single phoneme can force Wit to
mistranscribe the next seven words. In contrast, only two of the next 10 words
will be mistranscribed by the Google (Normal) model. This robustness for the
Google (Normal) model might be due to its internal recurrent layers being less
weighted towards the previously transcribed content. It is also interesting to
note that, despite their common internal structure, Deep Speech-1 and Deep
Speech-2 are significantly different in their vulnerability to this effect.
Deep Speech-1 and Deep Speech-2 have a cosine similarity score of 0.4 and 0.7,
respectively. This difference could potentially be attributed to different
feature extraction mechanisms. While Deep Speech-1 uses MFCCs, its counterpart
uses a CNN feature extraction. This is because feature extraction using MFCCs and CNNs produce varying results and might capture divergent information about the signal.


Observing the models individually, there is no observable relationship between
the cosine similarity and the phoneme type. All phonemes seem to be
approximately equally vulnerable to the attack. This means that models do not
use knowledge of previously occurring phonemes when transcribing the current
word. In fact, the results above show that the models use the current phonemes
and previous words in combination to transcribe the current word, which
intuitively is the intention behind modern ASR system design.




\vspace{-1em}
\begin{table}[tp]
\noindent\setlength\tabcolsep{2pt}
\begin{tabularx}{\textwidth}{lcccccc}
                                                                        & \multicolumn{1}{l}{}                                                           & \multicolumn{1}{l}{}                                                            & \textbf{To (g)}                                                                         & \multicolumn{1}{l}{}     & \multicolumn{1}{l}{}        & \multicolumn{1}{l}{}                                                           \\ \cline{2-7}
\multicolumn{1}{l|}{}                                                   & \multicolumn{1}{c|}{$P(S_{f \rightarrow g})$}                                                   & \multicolumn{1}{c|}{\begin{tabular}[c]{@{}c@{}}Google \\ (Phone)\end{tabular}} & \multicolumn{1}{c|}{\begin{tabular}[c]{@{}c@{}}Google\\ (Normal)\end{tabular}} & \multicolumn{1}{c|}{\begin{tabular}[c]{@{}c@{}}Wit\\\end{tabular}} & \multicolumn{1}{c|}{Sphinx} & \multicolumn{1}{c|}{\begin{tabular}[c]{@{}c@{}}Deep-Speech 1\end{tabular}} \\ \cline{2-7}
\multicolumn{1}{l|}{}                                                   & \multicolumn{1}{c|}{\begin{tabular}[c]{@{}c@{}}Google\\ (Phone)\end{tabular}}  & \multicolumn{1}{c|}{100\%}                                                          & \multicolumn{1}{c|}{\textbf{78\%}}                                                        & \multicolumn{1}{c|}{\textbf{83\%}}  & \multicolumn{1}{c|}{\textbf{42\%}}     & \multicolumn{1}{c|}{\textbf{87\%}}                                                          \\ \cline{2-7}
\multicolumn{1}{l|}{}    & \multicolumn{1}{c|}{\begin{tabular}[c]{@{}c@{}}Google\\ (Normal)\end{tabular}} & \multicolumn{1}{c|}{13\%}                                                         & \multicolumn{1}{c|}{100\%}                                                         & \multicolumn{1}{c|}{\textbf{65\%}}  & \multicolumn{1}{c|}{22\%}     & \multicolumn{1}{c|}{\textbf{70\%}}                                                          \\ \cline{2-7}
\multicolumn{1}{c|}{\begin{tabular}[c]{@{}c@{}}\textbf{From}\\ \textbf{(f)}\end{tabular}}                                                    & \multicolumn{1}{c|}{\begin{tabular}[c]{@{}c@{}}~\\Wit\end{tabular}}                                                       & \multicolumn{1}{c|}{6\%}                                                          & \multicolumn{1}{c|}{10\%}                                                        & \multicolumn{1}{c|}{100\%}   & \multicolumn{1}{c|}{14\%}     & \multicolumn{1}{c|}{\textbf{52\%}}                                                          \\ \cline{2-7}
\multicolumn{1}{l|}{}                                                   & \multicolumn{1}{c|}{\begin{tabular}[c]{@{}c@{}}~\\Sphinx\end{tabular}}                                                    & \multicolumn{1}{c|}{21\%}                                                         & \multicolumn{1}{c|}{\textbf{74\%}}                                                        & \multicolumn{1}{c|}{\textbf{81\%}}  & \multicolumn{1}{c|}{100\%}      & \multicolumn{1}{c|}{\textbf{80\%}}                                                          \\ \cline{2-7}
\multicolumn{1}{l|}{}                                                   & \multicolumn{1}{c|}{\begin{tabular}[c]{@{}c@{}}Deep-Speech 1\end{tabular}} & \multicolumn{1}{c|}{3\%}                                                          & \multicolumn{1}{c|}{7\%}                                                         & \multicolumn{1}{c|}{31\%}  & \multicolumn{1}{c|}{12\%}     & \multicolumn{1}{c|}{100\%}                                                         \\ \cline{2-7}
\end{tabularx}
\smallskip
\smallskip
\caption{The probability of transferability $P(S_{f \rightarrow g})$ calculated
for each combination of the tested models. Only `harder' models tend to
transfer well to weaker models. The elements in bold show the highest
transferability successes. Model names in the columns have been arranged in
descending order of their strength from \textit{harder} to \textit{weaker}.}
\label{tab:ssa_transferability_table}
\vspace{-1.6em}
\hrulefill
\vspace{-1.6em}
\end{table}

\begin{figure*}  
  \centering
  \includegraphics[width=\textwidth]{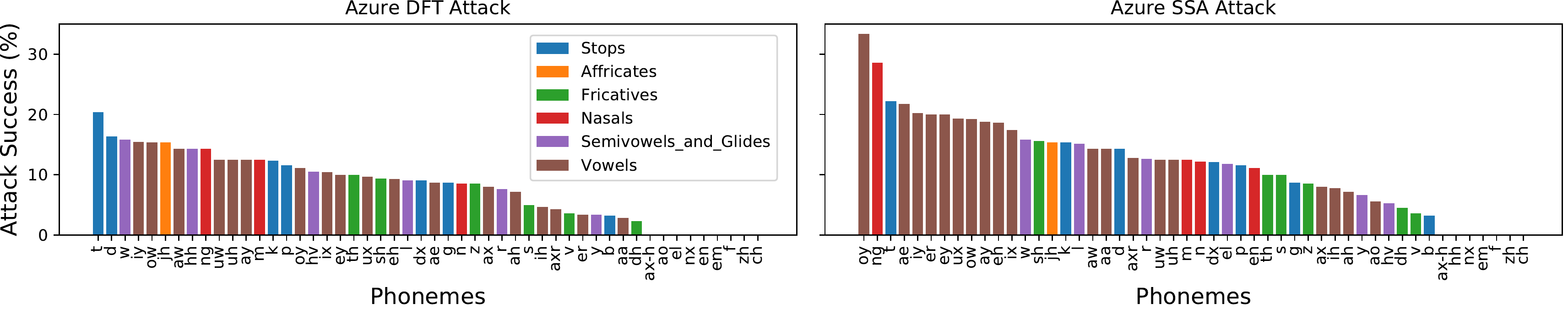}
  \caption{Success rate of our attack against a Automatic Voice Identification (AVI) system. When
      perturbing a single phoneme in the entire audio
  sample, an adversary has a greater chance of succeeding with
  an SSA attack rather than a DFT attack. Additionally, similar to the observation in Figure~\ref{fig:phoneme_atk_succ} vowels are more vulnerable than other phonemes.}
  \label{fig:Azure_phoneme_comp_akt_succ}
  \vspace{-1.5em}
  \hrulefill
  \vspace{-1.6em}
\end{figure*}

\subsection{Attacks Against AVI systems}
\label{sub:sr_attacks}
Next, we observe our attacks' effectiveness inducing errors an AVI system. Figure~\ref{fig:Azure_phoneme_comp_akt_succ} shows the attack success rate per vowel for both SSA and DFT attacks. Similar to our attack results against ASR models, attacks against the AVI system exhibited higher success rate when attacking phoneme. This means an adversary wishing to maximize attack success should focus on perturbing vowels.
Figure~\ref{fig:Azure_phoneme_comp_threshold} shows the relative amount of
perturbation necessary in order to force an AVI misclassification. The SSA
attack requires a relatively high perturbation for every
phoneme. In contrast, the DFT attack requires a smaller degree of perturbation, all except some vowels. This implies that
the DFT attack could be conducted more stealthily by intelligently selecting certain phonemes. The perturbation required, the more stealthy the attack. However,
because the SSA attack requires larger degree of perturbations for most phonemes, the level of stealth the attack can achieve is relatively higher.

\subsection{Transferability}\label{sub:Transferability}

Table~\ref{tab:ssa_transferability_table} shows results of the transferability
experiments using the SSA-based attack. Overall, the attack has the highest
transfer probability when a `harder' model is used to generate attack audio
samples. The Google (Phone) model had the highest average threshold across
samples which, as discussed in
Section~\ref{sub:transferability}, translated to the highest transfer probability. In
contrast, a weaker model will have a lower threshold and thus be less likely to
transfer. This can be seen when treating Sphinx
as the baseline model in Table~\ref{tab:ssa_transferability_table}. The table
shows that in the worst case attack audio generated for the Google (Phone)
the model will also be effective against any other model at least 42\% of the
time. This ensures a high probability of success even in the extreme case
when the adversary does not know which model the victim will be employing. By
generating attack samples on a stronger model, an adversary can expect an attack
to transfer well as long as the victim model is weaker. Finding a weaker
model is trivial. As long as the adversary has sufficient queries, they can compare
transcription rates for a candidate audio sample between the two models.

\subsection{Detection}\label{sub:Detection_results}
For the detection experiments, we created set of 266 adversarial samples, perturbed using the DFT technique, while the benign set consisted of 534 unperturbed audio samples. Of the adversarial samples provided to the Google Speech API, 20\% did \textit{not produce any transcription}. This was true both the entire audio samples and their corresponding partition. This means that the WER for these samples was 0, which introduced a bias to our results. Though this is perfect for the dissident, but it introduces bias in our results. Specifically, because there are two benign cases in which the WER will be zero, benign audio or very noisy audio. We discarded these audio samples from our adversarial set to remove this bias in our results. In the real world, an attacker can merely reduce the attack factor to prevent the model from producing no transcription. Next, we calculated the AUC scores for the samples. In our case, the AUC value was 0.527, which is far lower than the AUC value of 0.936 reported by the authors for the attacks they tested. This means even though the temporal based detection can do an excellent job of detecting other attacks, it is highly inaccurate for detecting our attack samples.








\begin{figure}
 \centering
  \includegraphics[width=0.8\columnwidth]{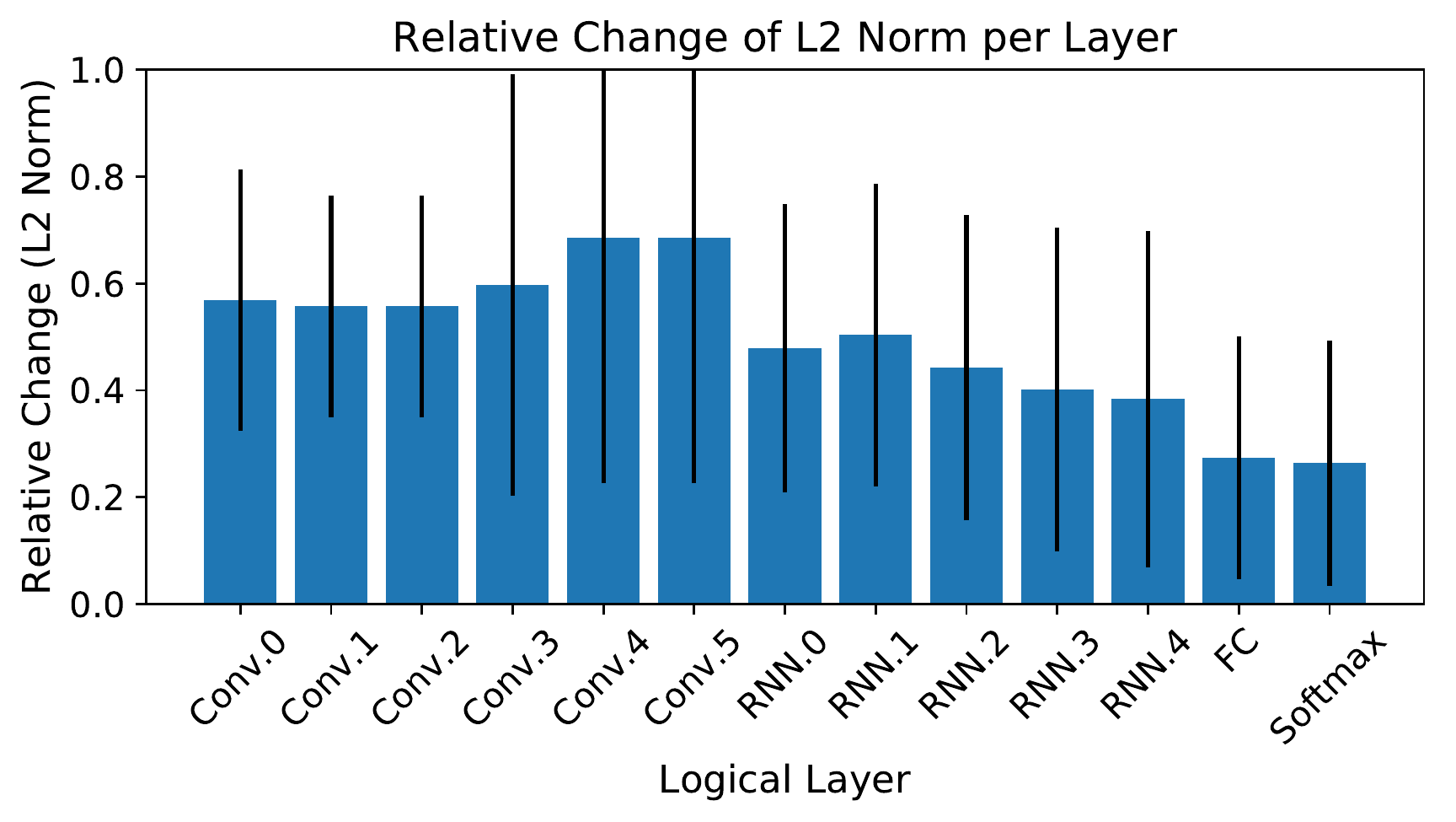}
  \caption{Calculated net change $\Delta h^l$ for each layer in the tested Deep Speech 2 model (Convolutional, Recurrent (RNN), fully connected (FC), and Softmax). Lower-level convolutional layers tend to have higher adversarial effect than upper-level recurrent and fully connected layers.}
  \label{fig:per_layer}
  \vspace{-1.6em}
  \hrulefill
\end{figure}

\subsection{Per-Layer Effects}\label{sub:per_layer_effects}

We now probe more deeply into why the attacks succeed.  In particular, we will
explore the attacks' effects at each layer of a typical deep learning-based
speech transcription model. We begin by opening our local deployment of Deep
Speech-2, which represents a state-of-the-art architecture in this task. The
model was provided by an open-source community and was not trained on the same
breadth of data as the original proposal~\cite{pmlr-DS2}. However, examining
such a deployment gives insight into the attack's effects on a publicly
available, open-source implementation. We begin by recording the activations at
each layer of the Deep Speech-2 architecture for several pairs of original and
word-level perturbed audio sample.

We treat the activation of the original audio as a baseline and subtract this
from the activation of the perturbed audio to find the net effect. We quantify
their distance using the $L_2$-norm, then divide this value by the $L_2$-norm
of the original activation. In total, the tested implementation is comprised
of thirteen logical layers\footnote{We treat the combination of convolutional,
pooling, and activation layers as one logical convolution layer. For recurrent
layers, we treat the BatchNorm-RNN~\cite{pmlr-DS2} as one logical recurrent
layer. }, thus for every normal audio $x$, adversarial audio $x^*$, layer
$h^l$, $l \in \{0,...,13\}$, and the $L_2$-norm written for brevity as $L_2$,
we have the layer's net change $\Delta h^l$ calculated using the equation 
	\(\Delta h^l = \frac{L_2[h^l(x) - h^l(x^*)]}{L_2[h^l(x)]}\).

This process was repeated for every benign-perturbed audio pair in our corpus, which is
the same corpus built for the word-level perturbation experiments in
Section~\ref{sub:word_level_perturbations}. We average across every pair, then
repeat this process for every logical layer $h^l$ in the Deep Speech-2
architecture to produce the results shown in Figure~\ref{fig:per_layer}. We
observe a higher rate of change for low-level convolutional layers,
particularly the upper-level convolutional layers.
These convolutional layers learn to mimic
and surpass the MFCC preprocessing filter. The net change immediately
diminishes as the sample passes through the recurrent layers, reflecting the
intuition that changes at individual frames, phonemes, or words may be
mitigated by temporal information. However, we also observe a non-zero net
change in the fully connected and softmax layers, indicating that the
shift was enough to force mistranscription. This points to the CNNs as the
main source of the attack vulnerability.

\subsection{Over-Cellular}\label{sub:over_cellular}
\begin{figure} 
    \centering
  \includegraphics[width=0.8\columnwidth]{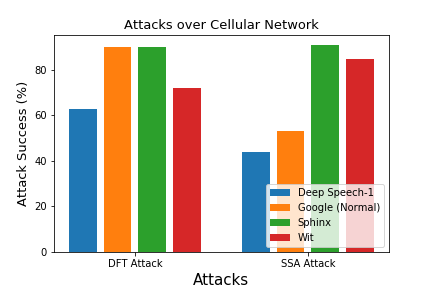}
  \caption{The attack audio was sent over the cellular network and recorded on a mobile end-point. The figure above shows the percentage of the attack audio that was still mistranscribed even after being passed over the cellular network.}
  \label{fig:attack_over_mobile_network}
  \vspace{-1.6em}
  \hrulefill
  \vspace{-1.6em}
\end{figure}

Next, we test our attacks for use over a cellular network. We
run previously successful attack audio samples over the cellular network before
passing them again to the target models. The rate of success for this
experiment is shown in Figure~\ref{fig:attack_over_mobile_network} plotted
against the DFT and SSA-based attacks for each model. If our  attack
were to be used over a cellular connection, having near real-time performance
is important. The largest source of potential delay caused by our
attack is from calculating the DFT on the original audio sample. While we do
not conduct our own time evaluation, Danielsson et al. showed that calculating a
DFT on a commodity Android platform took approximately 0.5 ms for a block
size of 4096 ~\cite{danielsson2017}.

The DFT-based attack managed to be more successful across the mobile network
than the SSA-based attack and was only consistently filtered by the Deep
Speech and Wit models. Overall, the models react differently based on
the transformation. Wit performs best under DFT transforms and worst
under SSA transforms, while the opposite is true for Deep Speech 1 and
Google (Normal) model. Sphinx is
equally vulnerable to both transformation methods.

An intuition for these results can be formed by considering the task each
transformation is performing. When transforming with the DFT, the attack audio
sample forms around pieces of weighted cross-correlations from the original
audio sample. In contrast, the SSA is built as a sum of interpretable
principal components. Although SSA may be able to produce fine-grained
perturbations to the signal, the granularity of perturbations is lost during
the mobile network's compression. This effect is expected to be amplified for
higher amounts of compression, although such a scenario also limits the
perceptibility of benign audio.

\subsection{Amazon Turk Listening Tests}\label{sub:amazon_turk_listening_tests}
In order to evaluate the transcription done by MTurk workers, we initially manually classified the
transcriptions as either correct or incorrect.
Table~\ref{tbl:mturk-transcription} shows a side by side comparison of
original and attack transcription of the perturbed portion of the audio
sample. Transcriptions which had either a missing word, a missing sentence, or
an additional word not present in the audio were marked as incorrect. At the
end of this classification task, the inter-rater agreement value, Cohen's
kappa was found to be $0.901$, which is generally considered to be `very good'
agreement among individual raters. Our manual evaluation found only $11\%$ of
the transcriptions to be incorrect. More specifically, we found that incorrect
transcriptions mostly had missing sentences from the beginning of the played
audio sample, but the transcriptions did not contain any misinterpreted words.
Our subjective evaluation did not consider wrong
transcription of perturbed vs. non-perturbed portion of the audio, rather we
only evaluated human comprehension of the audio sample.

In addition to subjective evaluation, we ran a phoneme-level edit distance test
on transcriptions to compare the level of transcription accuracy between perturbed
and non-perturbed audio samples. We used this formula for \textit{phoneme
edit distance, $\phi$}: $\phi =
\frac{\delta}{L}$, where $\delta$ is the Levenshtein edit distance between
phonemes of two transcriptions (original transcription and MTurk worker's
transcription) for a word and $L$ is the phoneme length of non-perturbed,
normal audio for the word~\cite{carlini2016hidden}. We defined accuracy as 1
when $\phi = 0$, indicating exact match between two transcriptions. For any
other value $\phi > 0$, we defined it as `in-accuracy' and assigned a value of
0. In Table~\ref{tbl:mturk}, we present transcription accuracy results between
perturbed and non-perturbed audio across our final sample size of 61. We also
ran a paired sample t-test, using the individual accuracy score for perturbed
and benign audio transcriptions, with the null hypothesis that participants'
accuracy levels were similar for both cases of transcriptions. Our results
showed participants had better accuracy transcribing non-perturbed audio
samples ($mean= 0.98, SD = 0.13$) than for perturbed audio ($mean = 0.90, SD =
0.30$). At a significance level of $p < 0.01$, our repeated-measures t-test
found this difference not to be significant, $t(60) = -2.315, p = 0.024$.
Recall that our chosen significance level ($p < 0.01$) was not arbitrary,
rather it was chosen during our sample size calculation for this study.
Together, this suggests that our word level audio perturbation create no
obstacle for human comprehension in telecommunication tasks, thus supporting
our null hypothesis.

\vspace{-1em}
\begin{table}[tp]
\centering
\begin{tabular}{|l|l|}
\hline

\textbf{Original Transcription}                                           & \textbf{Attack Transcription}                                               \\ \hline

\begin{tabular}[c]{@{}l@{}}How are you?\\ How's work going?\end{tabular}  & \begin{tabular}[c]{@{}l@{}}How are you\\ posmothdro?\end{tabular}           \\ \hline

\begin{tabular}[c]{@{}l@{}}I am really sorry\\ to hear that.\end{tabular} & \begin{tabular}[c]{@{}l@{}}I am relief for you\\ to hear that.\end{tabular} \\ \hline

\end{tabular}

\smallskip
\smallskip
\caption{Example of the attacked audio which was played to MTurk Workers and the
corresponding transcriptions. }
\label{tbl:mturk-transcription}
\vspace{-1.5em}
\hrulefill
\vspace{-1em}
\end{table}


\begin{table}[tp]
\centering
\begin{tabular}{cccc}
\multicolumn{2}{c}{\textbf{Accuracy (Perturbed)}}                                                                     & \multicolumn{2}{c}{\textbf{Accuracy (Benign)}}                                                                                                                           \\ \hline
\multicolumn{1}{|c|}{Male}                                                     & \multicolumn{1}{c|}{Female} & \multicolumn{1}{c|}{Male}                                                      & \multicolumn{1}{c|}{Female}                                                    \\ \hline
\multicolumn{1}{|c|}{\begin{tabular}[c]{@{}c@{}}91.8\%\\ (56/61)\end{tabular}} & \multicolumn{1}{c|}{\begin{tabular}[c]{@{}c@{}}100\%\\ (61/61)\end{tabular}}  & \multicolumn{1}{c|}{\begin{tabular}[c]{@{}c@{}}98.36\%\\ (60/61)\end{tabular}} & \multicolumn{1}{c|}{\begin{tabular}[c]{@{}c@{}}98.36\%\\ (60/61)
\end{tabular}} \\ \hline
\end{tabular}

\caption{Transcription accuracy results of MTurk workers for benign and perturbed audios between Male and Female speakers.}
\label{tbl:mturk}
\vspace{-1.5em}
\hrulefill
\vspace{-1.6em}
\end{table}

\section{Discussion}\label{sec:discussion}

\subsection{Phoneme vs. Word Level Perturbation}\label{para:disc_phon_vs_word}
Our attack aims to force a mistranscription while still being indistinguishable
from the original audio to a human listener. Our results indicate that at both
the phoneme-level and the word-level, the attack is able to fool black-box
models while keeping audio quality intact. However, the choice of performing
word-level or phoneme-level attacks is dependent on factors such as attack
success guarantee, audible distortion, and speed. The adversary can achieve
guaranteed attack success for any word in the dictionary if word-level
perturbations are used. However, this is not always true for a phoneme-level
perturbation, particularly for phonemes which are phonetically silent. An ASR
system may still properly transcribe the entire word even if the chosen phoneme
is maximally perturbed.  Phoneme-level perturbations may introduce less
human-audible distortion to the entire word, as the human brain is well suited
to interpolate speech and can compensate for a single perturbed phoneme. In
terms of speed, creating word-level perturbations is significantly slower than
creating phoneme-level perturbations. This is because a phoneme-level attack
requires perturbing only a fraction of the audio samples needed when attacking
an entire word.

\subsection{Steps to Maximize Attack Success}\label{para:disc_attack_checking}
An adversary wishing to launch an Over-Cellular evasion attack on an ASR system would
be best off using the DFT-based phoneme-level attack on vowels, as it guarantees
a high level of attack success. Our transferability results show that an
attacker can generate samples for a known `hard' model such as Google (Phone)
and have reasonable confidence that the attack audio will transfer to an unknown
ASR model. From our ASR poisoning results, we observe that an adversary
does not have to perturb every word to earn 100\% mistranscription of the
utterance. \textit{Instead, the attacker can perturb a vowel of every other word
in the worst case, and every fifth word in the best case.} The ASR
poisoning effect will ensure that the non-perturbed words are also
mistranscribed. Finally, the attack audio samples have a high probability of
surviving the compression of a cellular network, which will enable the success
of our attack over lossy and noisy mediums.

Contrary to an ASR system attack, an adversary looking to execute an evasion
attack on an AVI system would prefer to use the SSA-based phoneme-level attack.
Similar to ASR poisoning, we observe that an adversary does not have to
perturb the entire sentence to cause a misidentification, but rather just a
single phoneme of a word in the sentence. \textit{Based on our results, the
attacker would need to perturb on average one phoneme every 8 words (33 phoneme)
to ensure a high likelihood of attack success.} The attack audio samples are
generated in an identical manner for both the ASR and AVI system attacks, thus
the AVI attack audio should also be robust against lossy and noisy mediums (e.g.,
a cellular network).

\subsection{Why the Attack Works}\label{para:attack_success_explanation} Our
attacks exploit the fundamental difference in how the human brain and ASR/AVI
systems process speech. Specifically, our attack discards low intensity
components of an audio sample which the human brain is primed to ignore. The
remaining components are enough for a human listener to correctly interpret the
perturbed audio sample. On the other hand, the ASR or AVI systems have unintentionally
learned to depend on these low intensity components for inference. This explains
why removing such insignificant parts of speech confuses the model and causes a
mistranscription or misidentification. Additionally, this may also explain some 
portion of the ASR and AVI error on regular testing data sets. Future work may use these 
revelations in order to build more robust models and be able to explain and reduce ASR 
and AVI system error.

\subsection{Audio CAPTCHAs}
\label{para:apps}
In addition to helping dissidents overcome mass-surveillance, our attack has other applications as well. Specifically, in the domain of audio
CAPTCHAs. These are often used by web services to validate the presence of a human.
CAPTCHAs relies on humans being able to transcribe audio better than machines,
an assumption that modern ASR systems call into
question~\cite{bursztein2011failure, tam2009breaking, sano2013solving,
solanki2017cyber, bock2017uncaptcha}.  Our attack could potentially be used to
intelligently distort audio CAPTCHAs as a countermeasure to modern ASR systems.

\section{Related Work}
\label{sec:related}
Machine Learning (ML) models, and in particular deep learning models, have shown
great performance advancements in previously complex tasks, such as image
classification~\cite{imagenet_krizhevsky12,inception_net15,resnet_He15} and
speech recognition~\cite{kaldi_hmm,pmlr-DS2,graves2014towards}. However,
previous work has shown that ML models are inherently vulnerable to a class of
ML known as Adversarial Machine Learning (AML)~\cite{Huang2011}.

Early AML techniques focused on visually imperceptible changes to an image
that cause the model to incorrectly classify the image. Such
attacks target either specific pixels~\cite{kurakin2016adversarial,
    szegedy2013intriguing, goodfellow2014explaining, baluja2017adversarial,
su2017one, moosavi2016deepfool}, or entire patches of
pixels~\cite{brown2017adversarial, sharif2016accessorize,
papernot2016limitations, carlini2017towards}. In some cases, the attack
generates entirely new images that the model would classify to an adversary's
chosen target~\cite{nguyen2015deep,liu2017trojaning}.

However, the success of these attacks are a result of two restrictive
assumptions. First, the attacks assume that the underlying target model is a
form of a neural network. Second, they assume the model can be influenced by
changes at the pixel level~\cite{papernot2016limitations, nguyen2015deep}. These
assumptions prevent image attacks from being used against ASR models. 
ASR systems have found success across a variety of ML
architectures, from Hidden Markov Models (HMMs) to Deep Neural Networks (DNNs).
Further, since audio data is normally preprocessed for feature extraction before
entering the statistical model, the models initially operate at a higher level
than the `pixel level' of their image counterparts. 



To overcome these limitations, previous works have proposed several new attacks
that exploit behaviors of particular models. These attacks can be categorized
into three broad techniques that generate audio that include:
\begin{inlineenum}  
\item \emph{inaudible} to the human ear but will be detected by the speech
    recognition model~\cite{zhang2017dolphinattack}, \item \emph{noisy} such that
    it
might sound like noise to the human, but will be correctly deciphered by the
automatic speech recognition~\cite{vaidya2015cocaine, carlini2016hidden,
biometrics}, and 
\item \emph{pristine} audio such that the audio sounds normal to the human but
    will be deciphered to a different, chosen
    phrase~\cite{yuan2018commandersong,cai2018attacking,gong2017crafting,kereliuk2015deep,
    alzantot2018did,kreuk2018fooling,cisse2017houdini,schonherr2018adversarial,erichennenfentskill}.
\end{inlineenum}
Although they may seem the most useful, attacks in the third category are
limited in their success, as they often require white-box access to the model.

Attacks against image recognition models are well studied, giving attackers
the ability to execute targeted attacks even in black-box settings. This has not yet
been possible against speech models~\cite{bispham2018taxonomy}, even for
untargeted attacks in a query efficient manner. That is, both targeted and
untargeted attacks require knowledge of the model internals (such as
architecture and parameterization) and large number of queries to the
model. In contrast, we propose a query efficient black-box attack
that is able to generate an attack audio sample that will be reliably mistranscribed by
the model, regardless of architecture or parameterization. Our attack can
generate an attack audio sample in logarithmic time, while leaving the audio
quality mostly unaffected.

\section{Conclusion}
\label{sec:conclusion}

Automatic speech recognition systems are playing an increasingly important role
in security decisions. As such, the robustness of these systems (and the
foundations upon which they are built) must be rigorously evaluated. We perform
such an evaluation in this paper, with particular focus on
speech-transcription. By exhibiting black-box attacks against of multiple
models, we demonstrate that such systems rely on audio features which are not
critical to human comprehension and are therefore vulnerable to
mistranscription attacks when such features are removed.  We then show that
such attacks can be efficiently conducted as perturbations to certain phonemes
(e.g., vowels) that cause significantly greater misclassification to the words that
follow them.  Finally, we not only demonstrate that our attacks can work across
models, but also show that the audio generated has no impact on
understandability to users. This detail is critical, as attacks that simply
obscure audio and make it useless to everyone are not particularly useful to
the adversaries we consider. While adversarial training may help in partial
mitigations, we believe that more substantial defenses are ultimately required to
defend against these attacks.

%
\bibliographystyle{IEEEtranS}
\bibliography{transcription_evasion}

\begin{thebibliography}{10}
\providecommand{\url}[1]{#1}
\csname url@samestyle\endcsname
\providecommand{\newblock}{\relax}
\providecommand{\bibinfo}[2]{#2}
\providecommand{\BIBentrySTDinterwordspacing}{\spaceskip=0pt\relax}
\providecommand{\BIBentryALTinterwordstretchfactor}{4}
\providecommand{\BIBentryALTinterwordspacing}{\spaceskip=\fontdimen2\font plus
\BIBentryALTinterwordstretchfactor\fontdimen3\font minus
  \fontdimen4\font\relax}
\providecommand{\BIBforeignlanguage}[2]{{%
\expandafter\ifx\csname l@#1\endcsname\relax
\typeout{** WARNING: IEEEtranS.bst: No hyphenation pattern has been}%
\typeout{** loaded for the language `#1'. Using the pattern for}%
\typeout{** the default language instead.}%
\else
\language=\csname l@#1\endcsname
\fi
#2}}
\providecommand{\BIBdecl}{\relax}
\BIBdecl

\bibitem{1k_words}
``{1,000 Most Common {US} English Words},'' Last accessed in 2019, available at
  \url{https://www.ef.edu/english-resources/english-vocabulary/top-1000-words/}.

\bibitem{azure_attest}
``Azure speaker identification api,'' Last accessed in 2019, available at
  \url{https://azure.microsoft.com/en-us/services/cognitive-servic/speaker-recognition/}.

\bibitem{usa_phone_calls}
``{Background on CTIA’s Semi-Annual Wireless Industry Survey },'' Last
  accessed in 2019, available at
  \url{http://files.ctia.org/pdf/CTIA_Survey_YE_2012_Graphics-FINAL.pdf}.

\bibitem{google_normal}
``{Google Cloud Speech-to-Text API},'' Last accessed in 2019, available at
  \url{https://cloud.google.com/speech-to-text/}.

\bibitem{google_speech_error}
``{Googles Speech Recognition Technology Now Has a 4.9\% Word Error Rate},''
  Last accessed in 2019, available at \url{https://bit.ly/2rGRtUQ}.

\bibitem{inside-chinas-massive-surveillance-operation}
``{Inside China's Massive Surveillance Operation},'' Last accessed in 2019,
  available at
  \url{https://www.wired.com/story/inside-chinas-massive-surveillance-operation/}.

\bibitem{mozilla_ds}
``Mozilla project deepspeech,'' Last accessed in 2019, available at
  \url{https://azure.microsoft.com/en-us/services/cognitive-servic/speaker-recognition/}.

\bibitem{nsa-speech-recognition-snowden-searchable-text}
``{NSA Speech Recognition Snowden Searchable Text},'' Last accessed in 2019,
  available at
  \url{https://theintercept.com/2015/05/05/nsa-speech-recognition-snowden-searchable-text/}.

\bibitem{shotooka}
``{Project {SHTOOKA} - A Multilingual Database of Audio Recordings of Words and
  Sentences},'' Last accessed in 2019, available at \url{http://shtooka.net/}.

\bibitem{tf_speech_rec}
``{Simple Audio Recognition},'' Last accessed in 2019, available at
  \url{https://www.tensorflow.org/tutorials/sequences/audio_recognition}.

\bibitem{an4_db}
``{The CMU Audio Database (also known as AN4 database)},'' Last accessed in
  2019, available at \url{http://www.speech.cs.cmu.edu/databases/an4/}.

\bibitem{google_phone}
``{Transcribing Phone Audio with Enhanced Models},'' Last accessed in 2019,
  available at \url{https://cloud.google.com/speech-to-text/docs/phone-model}.

\bibitem{twilio}
``{Twilio - Communication APIs for SMS, Voice, Video and Authentication},''
  Last accessed in 2019, available at \url{https://www.twilio.com/}.

\bibitem{wer_are_we}
``Wer are we - an attempt at tracking states of the art(s) and recent results
  on speech recognition,'' \url{https://github.com/syhw/wer_are_we}, Last
  accessed in 2019.

\bibitem{article_acc_models}
``{Who's Smartest: Alexa, Siri, and or Google Now?}'' Last accessed in 2019,
  available at \url{https://bit.ly/2ScTpX7}.

\bibitem{wit}
``{Wit.ai Natural Language for Developers},'' Last accessed in 2019, available
  at \url{https://wit.ai/}.

\bibitem{ICASSP12}
O.~Abdel-Hamid, A.-r. Mohamed, H.~Jiang, and G.~Penn, ``Applying convolutional
  neural networks concepts to hybrid nn-hmm model for speech recognition,'' pp.
  4277--4280, 05 2012.

\bibitem{biometrics}
H.~Abdullah, W.~Garcia, C.~Peeters, P.~Traynor, K.~Butler, and J.~Wilson,
  ``Practical hidden voice attacks against speech and speaker recognition
  systems,'' \emph{Proceedings of the 2019 Network and Distributed System
  Security Symposium (NDSS)}, 2019.

\bibitem{alzantot2018did}
M.~Alzantot, B.~Balaji, and M.~Srivastava, ``Did you hear that? adversarial
  examples against automatic speech recognition,'' \emph{arXiv preprint
  arXiv:1801.00554}, 2018.

\bibitem{pmlr-DS2}
\BIBentryALTinterwordspacing
D.~Amodei~et al., ``Deep speech 2 : End-to-end speech recognition in english
  and mandarin,'' in \emph{Proceedings of The 33rd International Conference on
  Machine Learning}, ser. Proceedings of Machine Learning Research, M.~F.
  Balcan and K.~Q. Weinberger, Eds., vol.~48.\hskip 1em plus 0.5em minus
  0.4em\relax New York, New York, USA: PMLR, 20--22 Jun 2016, pp. 173--182.
  [Online]. Available: \url{http://proceedings.mlr.press/v48/amodei16.html}
\BIBentrySTDinterwordspacing

\bibitem{bahdanau2016end}
D.~Bahdanau, J.~Chorowski, D.~Serdyuk, P.~Brakel, and Y.~Bengio, ``End-to-end
  attention-based large vocabulary speech recognition,'' in \emph{Acoustics,
  Speech and Signal Processing (ICASSP), 2016 IEEE International Conference
  on}.\hskip 1em plus 0.5em minus 0.4em\relax IEEE, 2016, pp. 4945--4949.

\bibitem{baluja2017adversarial}
S.~Baluja and I.~Fischer, ``Adversarial transformation networks: Learning to
  generate adversarial examples,'' \emph{arXiv preprint arXiv:1703.09387},
  2017.

\bibitem{bispham2018taxonomy}
M.~K. Bispham, I.~Agrafiotis, and M.~Goldsmith, ``{A Taxonomy of Attacks via
  the Speech Interface},'' 2018.

\bibitem{bock2017uncaptcha}
K.~Bock, D.~Patel, G.~Hughey, and D.~Levin, ``uncaptcha: a low-resource defeat
  of recaptcha's audio challenge,'' in \emph{Proceedings of the 11th USENIX
  Conference on Offensive Technologies}.\hskip 1em plus 0.5em minus 0.4em\relax
  USENIX Association, 2017, pp. 7--7.

\bibitem{brown2017adversarial}
T.~B. Brown, D.~Man{\'e}, A.~Roy, M.~Abadi, and J.~Gilmer, ``Adversarial
  patch,'' \emph{arXiv preprint arXiv:1712.09665}, 2017.

\bibitem{bursztein2011failure}
E.~Bursztein, R.~Beauxis, H.~Paskov, D.~Perito, C.~Fabry, and J.~Mitchell,
  ``The failure of noise-based non-continuous audio captchas,'' in
  \emph{Security and Privacy (SP), 2011 IEEE Symposium on}.\hskip 1em plus
  0.5em minus 0.4em\relax IEEE, 2011, pp. 19--31.

\bibitem{cai2018attacking}
W.~Cai, A.~Doshi, and R.~Valle, ``Attacking speaker recognition with deep
  generative models,'' \emph{arXiv preprint arXiv:1801.02384}, 2018.

\bibitem{carlini2016hidden}
N.~Carlini, P.~Mishra, T.~Vaidya, Y.~Zhang, M.~Sherr, C.~Shields, D.~Wagner,
  and W.~Zhou, ``Hidden voice commands.'' in \emph{USENIX Security Symposium},
  2016, pp. 513--530.

\bibitem{carlini2017towards}
N.~Carlini and D.~Wagner, ``Towards evaluating the robustness of neural
  networks,'' in \emph{Security and Privacy (SP), 2017 IEEE Symposium
  on}.\hskip 1em plus 0.5em minus 0.4em\relax IEEE, 2017, pp. 39--57.

\bibitem{carliniL2}
N.~{Carlini} and D.~{Wagner}, ``{Audio Adversarial Examples: Targeted Attacks
  on Speech-to-Text},'' \emph{ArXiv e-prints}, p. arXiv:1801.01944, Jan. 2018.

\bibitem{chorowski2015attention}
J.~K. Chorowski, D.~Bahdanau, D.~Serdyuk, K.~Cho, and Y.~Bengio,
  ``Attention-based models for speech recognition,'' in \emph{Advances in
  neural information processing systems}, 2015, pp. 577--585.

\bibitem{cisse2017houdini}
M.~Cisse, Y.~Adi, N.~Neverova, and J.~Keshet, ``Houdini: Fooling deep
  structured prediction models,'' \emph{arXiv preprint arXiv:1707.05373}, 2017.

\bibitem{danielsson2017}
\BIBentryALTinterwordspacing
A.~Danielsson, ``Comparing android runtime with native: Fast fourier transform
  on android,'' 2017, mS thesis. [Online]. Available:
  \url{"https://bit.ly/2MQpUV1"}
\BIBentrySTDinterwordspacing

\bibitem{dohmatob2018limitations}
E.~Dohmatob, ``Limitations of adversarial robustness: strong no free lunch
  theorem,'' \emph{arXiv preprint arXiv:1810.04065}, 2018.

\bibitem{garofolo1988getting}
J.~S. Garofolo \emph{et~al.}, ``Getting started with the darpa timit cd-rom: An
  acoustic phonetic continuous speech database,'' \emph{National Institute of
  Standards and Technology (NIST), Gaithersburgh, MD}, vol. 107, p.~16, 1988.

\bibitem{masking_book}
S.~A. Gelfand, \emph{Hearing: An Introduction to Psychological and
  Physiological Acoustics}, 5th~ed.\hskip 1em plus 0.5em minus 0.4em\relax
  Informa Healthcare, 2009.

\bibitem{gong2017crafting}
Y.~Gong and C.~Poellabauer, ``Crafting adversarial examples for speech
  paralinguistics applications,'' \emph{arXiv preprint arXiv:1711.03280}, 2017.

\bibitem{goodfellow2014explaining}
I.~J. Goodfellow, J.~Shlens, and C.~Szegedy, ``Explaining and harnessing
  adversarial examples,'' \emph{arXiv preprint arXiv:1412.6572}, 2014.

\bibitem{graves2014towards}
A.~Graves and N.~Jaitly, ``Towards end-to-end speech recognition with recurrent
  neural networks,'' in \emph{International Conference on Machine Learning},
  2014, pp. 1764--1772.

\bibitem{graves13}
A.~Graves, A.-r. Mohamed, and G.~Hinton, ``Speech recognition with deep
  recurrent neural networks,'' \emph{ICASSP, IEEE International Conference on
  Acoustics, Speech and Signal Processing - Proceedings}, vol.~38, 03 2013.

\bibitem{hannun2014deep}
A.~Hannun, C.~Case, J.~Casper, B.~Catanzaro, G.~Diamos, E.~Elsen, R.~Prenger,
  S.~Satheesh, S.~Sengupta, A.~Coates \emph{et~al.}, ``Deep speech: Scaling up
  end-to-end speech recognition,'' \emph{arXiv preprint arXiv:1412.5567}, 2014.

\bibitem{resnet_He15}
K.~He, X.~Zhang, S.~Ren, and J.~Sun, ``{Deep Residual Learning for Image
  Recognition},'' in \emph{Proceedings of the IEEE conference on computer
  vision and pattern recognition}, 2016, pp. 770--778.

\bibitem{Huang2011}
\BIBentryALTinterwordspacing
L.~Huang, A.~D. Joseph, B.~Nelson, B.~I. Rubinstein, and J.~D. Tygar,
  ``Adversarial machine learning,'' in \emph{Proceedings of the 4th ACM
  Workshop on Security and Artificial Intelligence}, ser. AISec '11.\hskip 1em
  plus 0.5em minus 0.4em\relax New York, NY, USA: ACM, 2011, pp. 43--58.
  [Online]. Available: \url{http://doi.acm.org/10.1145/2046684.2046692}
\BIBentrySTDinterwordspacing

\bibitem{kereliuk2015deep}
C.~Kereliuk, B.~L. Sturm, and J.~Larsen, ``Deep learning and music
  adversaries,'' \emph{IEEE Transactions on Multimedia}, vol.~17, no.~11, pp.
  2059--2071, 2015.

\bibitem{kopcke2010evaluation}
H.~K{\"o}pcke, A.~Thor, and E.~Rahm, ``Evaluation of entity resolution
  approaches on real-world match problems,'' \emph{Proceedings of the VLDB
  Endowment}, vol.~3, no. 1-2, pp. 484--493, 2010.

\bibitem{kreuk2018fooling}
F.~Kreuk, Y.~Adi, M.~Cisse, and J.~Keshet, ``Fooling end-to-end speaker
  verification by adversarial examples,'' \emph{arXiv preprint
  arXiv:1801.03339}, 2018.

\bibitem{imagenet_krizhevsky12}
A.~Krizhevsky, I.~Sutskever, and G.~E.~Hinton, ``Imagenet classification with
  deep convolutional neural networks,'' \emph{Neural Information Processing
  Systems}, vol.~25, 01 2012.

\bibitem{erichennenfentskill}
D.~Kumar, R.~Paccagnella, P.~Murley, E.~Hennenfent, J.~Mason, A.~Bates, and
  M.~Bailey, ``Skill squatting attacks on amazon alexa,'' in \emph{27th
  {USENIX} Security Symposium ({USENIX} Security 18)}.\hskip 1em plus 0.5em
  minus 0.4em\relax {USENIX} Association, 2018.

\bibitem{kurakin2016adversarial}
A.~Kurakin, I.~Goodfellow, and S.~Bengio, ``Adversarial examples in the
  physical world,'' \emph{arXiv preprint arXiv:1607.02533}, 2016.

\bibitem{sphinx}
P.~Lamere, P.~Kwok, W.~Walker, E.~Gouv\^{e}a, R.~Singh, B.~Raj, and P.~Wolf,
  ``Design of the cmu sphinx-4 decoder,'' in \emph{Eighth European Conference
  on Speech Communication and Technology}, 2003.

\bibitem{liu2017trojaning}
Y.~Liu, S.~Ma, Y.~Aafer, W.-C. Lee, J.~Zhai, W.~Wang, and X.~Zhang, ``Trojaning
  attack on neural networks,'' \emph{Proceedings of the 2017 Network and
  Distributed System Security Symposium (NDSS)}, 2017.

\bibitem{2017arXiv170703501L}
J.~{Lu}, H.~{Sibai}, E.~{Fabry}, and D.~{Forsyth}, ``{NO Need to Worry about
  Adversarial Examples in Object Detection in Autonomous Vehicles},''
  \emph{ArXiv e-prints}, 2017.

\bibitem{madry2017towards}
A.~Madry, A.~Makelov, L.~Schmidt, D.~Tsipras, and A.~Vladu, ``Towards deep
  learning models resistant to adversarial attacks,'' \emph{arXiv preprint
  arXiv:1706.06083}, 2017.

\bibitem{moosavi2016deepfool}
S.~M. Moosavi~Dezfooli, A.~Fawzi, and P.~Frossard, ``Deepfool: a simple and
  accurate method to fool deep neural networks,'' in \emph{Proceedings of 2016
  IEEE Conference on Computer Vision and Pattern Recognition (CVPR)}, no.
  EPFL-CONF-218057, 2016.

\bibitem{sainath13}
T.~N.~Sainath, A.-r. Mohamed, B.~Kingsbury, and B.~Ramabhadran, ``Deep
  convolutional neural networks for lvcsr,'' pp. 8614--8618, 05 2013.

\bibitem{ds2_pytorch}
S.~Naren, ``Speech recognition using deepspeech-2,'' Last accessed in 2019,
  available at \url{https://github.com/SeanNaren/deepspeech.pytorch}.

\bibitem{nguyen2015deep}
A.~Nguyen, J.~Yosinski, and J.~Clune, ``Deep neural networks are easily fooled:
  High confidence predictions for unrecognizable images,'' in \emph{Proceedings
  of the IEEE Conference on Computer Vision and Pattern Recognition}, 2015, pp.
  427--436.

\bibitem{panayotov2015librispeech}
V.~Panayotov, G.~Chen, D.~Povey, and S.~Khudanpur, ``Librispeech: an asr corpus
  based on public domain audio books,'' in \emph{Acoustics, Speech and Signal
  Processing (ICASSP), 2015 IEEE International Conference on}.\hskip 1em plus
  0.5em minus 0.4em\relax IEEE, 2015, pp. 5206--5210.

\bibitem{papernot17}
N.~Papernot, P.~McDaniel, I.~Goodfellow, S.~Jha, Z.~B. Celik, and A.~Swami,
  ``{Practical Black-box Attacks Against Machine Learning},'' in
  \emph{Proceedings of the 2017 ACM on Asia Conference on Computer and
  Communications Security}.\hskip 1em plus 0.5em minus 0.4em\relax ACM, 2017,
  pp. 506--519.

\bibitem{papernot2016limitations}
N.~Papernot, P.~McDaniel, S.~Jha, M.~Fredrikson, Z.~B. Celik, and A.~Swami,
  ``The limitations of deep learning in adversarial settings,'' in
  \emph{Security and Privacy (EuroS\&P), 2016 IEEE European Symposium
  on}.\hskip 1em plus 0.5em minus 0.4em\relax IEEE, 2016, pp. 372--387.

\bibitem{kaldi_hmm}
D.~Povey, A.~Ghoshal, G.~Boulianne, L.~Burget, O.~Glembek, N.~Goel,
  M.~Hannemann, P.~Motl\'{i}\v{c}ek, Y.~Qian, P.~Schwarz, J.~Silovsk\'{y},
  G.~Stemmer, and K.~Vesel\'{y}, ``The kaldi speech recognition toolkit,'' in
  \emph{IEEE 2011 Workshop on Automatic Speech Recognition and
  Understanding}.\hskip 1em plus 0.5em minus 0.4em\relax IEEE Signal Processing
  Society, 2011, iEEE Catalog No.: CFP11SRW-USB.

\bibitem{qin2019imperceptible}
Y.~Qin, N.~Carlini, I.~Goodfellow, G.~Cottrell, and C.~Raffel, ``Imperceptible,
  robust, and targeted adversarial examples for automatic speech recognition,''
  \emph{arXiv preprint arXiv:1903.10346}, 2019.

\bibitem{rabiner1993fundamentals}
L.~R. Rabiner and B.-H. Juang, \emph{Fundamentals of speech recognition}.\hskip
  1em plus 0.5em minus 0.4em\relax PTR Prentice Hall Englewood Cliffs, 1993,
  vol.~14.

\bibitem{rabiner1978digital}
L.~R. Rabiner and R.~W. Schafer, \emph{Digital processing of speech
  signals}.\hskip 1em plus 0.5em minus 0.4em\relax Prentice Hall, 1978.

\bibitem{rix2001perceptual}
A.~W. Rix, J.~G. Beerends, M.~P. Hollier, and A.~P. Hekstra, ``Perceptual
  evaluation of speech quality (pesq)-a new method for speech quality
  assessment of telephone networks and codecs,'' in \emph{2001 IEEE
  International Conference on Acoustics, Speech, and Signal Processing.
  Proceedings (Cat. No. 01CH37221)}, vol.~2.\hskip 1em plus 0.5em minus
  0.4em\relax IEEE, 2001, pp. 749--752.

\bibitem{sainath15}
T.~N. Sainath, O.~Vinyals, A.~Senior, and H.~Sak, ``Convolutional, long
  short-term memory, fully connected deep neural networks,'' in \emph{2015 IEEE
  International Conference on Acoustics, Speech and Signal Processing
  (ICASSP)}, April 2015, pp. 4580--4584.

\bibitem{sak14}
H.~Sak, A.~Senior, and F.~Beaufays, ``Long short-term memory recurrent neural
  network architectures for large scale acoustic modeling,'' \emph{Proceedings
  of the Annual Conference of the International Speech Communication
  Association, INTERSPEECH}, pp. 338--342, 01 2014.

\bibitem{sak_vinalys14}
H.~Sak, O.~Vinyals, G.~Heigold, A.~Senior, E.~McDermott, R.~Monga, and M.~Mao,
  ``Sequence discriminative distributed training of long short-term memory
  recurrent neural networks,'' \emph{Proceedings of the Annual Conference of
  the International Speech Communication Association, INTERSPEECH}, pp.
  1209--1213, 01 2014.

\bibitem{sak15}
\BIBentryALTinterwordspacing
H.~Sak, A.~Senior, K.~Rao, and F.~Beaufays, ``Fast and accurate recurrent
  neural network acoustic models for speech recognition,'' \emph{CoRR}, vol.
  abs/1507.06947, 2015. [Online]. Available:
  \url{http://arxiv.org/abs/1507.06947}
\BIBentrySTDinterwordspacing

\bibitem{sano2013solving}
S.~Sano, T.~Otsuka, and H.~G. Okuno, ``Solving google’s continuous audio
  captcha with hmm-based automatic speech recognition,'' in \emph{International
  Workshop on Security}.\hskip 1em plus 0.5em minus 0.4em\relax Springer, 2013,
  pp. 36--52.

\bibitem{schonherr2018adversarial}
L.~Sch{\"o}nherr, K.~Kohls, S.~Zeiler, T.~Holz, and D.~Kolossa, ``Adversarial
  attacks against automatic speech recognition systems via psychoacoustic
  hiding,'' \emph{arXiv preprint arXiv:1808.05665}, 2018.

\bibitem{sharif2016accessorize}
M.~Sharif, S.~Bhagavatula, L.~Bauer, and M.~K. Reiter, ``Accessorize to a
  crime: Real and stealthy attacks on state-of-the-art face recognition,'' in
  \emph{Proceedings of the 2016 ACM SIGSAC Conference on Computer and
  Communications Security}.\hskip 1em plus 0.5em minus 0.4em\relax ACM, 2016,
  pp. 1528--1540.

\bibitem{solanki2017cyber}
S.~Solanki, G.~Krishnan, V.~Sampath, and J.~Polakis, ``In (cyber) space bots
  can hear you speak: Breaking audio captchas using ots speech recognition,''
  in \emph{Proceedings of the 10th ACM Workshop on Artificial Intelligence and
  Security}.\hskip 1em plus 0.5em minus 0.4em\relax ACM, 2017, pp. 69--80.

\bibitem{su2017one}
J.~Su, D.~V. Vargas, and S.~Kouichi, ``One pixel attack for fooling deep neural
  networks,'' \emph{arXiv preprint arXiv:1710.08864}, 2017.

\bibitem{inception_net15}
C.~Szegedy, W.~Liu, Y.~Jia, P.~Sermanet, S.~Reed, D.~Anguelov, D.~Erhan,
  V.~Vanhoucke, and A.~Rabinovich, ``{Going Deeper with Convolutions},'' in
  \emph{2015 IEEE Conference on Computer Vision and Pattern Recognition
  (CVPR)}, June 2015, pp. 1--9.

\bibitem{szegedy2013intriguing}
C.~Szegedy, W.~Zaremba, I.~Sutskever, J.~Bruna, D.~Erhan, I.~Goodfellow, and
  R.~Fergus, ``Intriguing properties of neural networks,'' \emph{arXiv preprint
  arXiv:1312.6199}, 2013.

\bibitem{tam2009breaking}
J.~Tam, J.~Simsa, S.~Hyde, and L.~V. Ahn, ``Breaking audio captchas,'' in
  \emph{Advances in Neural Information Processing Systems}, 2009, pp.
  1625--1632.

\bibitem{taori2018targeted}
R.~Taori, A.~Kamsetty, B.~Chu, and N.~Vemuri, ``Targeted adversarial examples
  for black box audio systems,'' \emph{arXiv preprint arXiv:1805.07820}, 2018.

\bibitem{tsipras2018robustness}
D.~Tsipras, S.~Santurkar, L.~Engstrom, A.~Turner, and A.~Madry, ``Robustness
  may be at odds with accuracy,'' \emph{arXiv preprint arXiv:1805.12152},
  vol.~1, 2018.

\bibitem{vaidya2015cocaine}
T.~Vaidya, Y.~Zhang, M.~Sherr, and C.~Shields, ``Cocaine noodles: exploiting
  the gap between human and machine speech recognition,'' \emph{WOOT}, vol.~15,
  pp. 10--11, 2015.

\bibitem{venugopalan2014translating}
S.~Venugopalan, H.~Xu, J.~Donahue, M.~Rohrbach, R.~Mooney, and K.~Saenko,
  ``Translating videos to natural language using deep recurrent neural
  networks,'' \emph{arXiv preprint arXiv:1412.4729}, 2014.

\bibitem{yang2018characterizing}
Z.~Yang, B.~Li, P.-Y. Chen, and D.~Song, ``Characterizing audio adversarial
  examples using temporal dependency,'' \emph{arXiv preprint arXiv:1809.10875},
  2018.

\bibitem{yuan2018commandersong}
X.~Yuan, Y.~Chen, Y.~Zhao, Y.~Long, X.~Liu, K.~Chen, S.~Zhang, H.~Huang,
  X.~Wang, and C.~A. Gunter, ``Commandersong: A systematic approach for
  practical adversarial voice recognition,'' in \emph{{Proceedings of the
  USENIX Security Symposium}}, 2018.

\bibitem{zhang2017dolphinattack}
G.~Zhang, C.~Yan, X.~Ji, T.~Zhang, T.~Zhang, and W.~Xu, ``Dolphinattack:
  Inaudible voice commands,'' in \emph{Proceedings of the 2017 ACM SIGSAC
  Conference on Computer and Communications Security}.\hskip 1em plus 0.5em
  minus 0.4em\relax ACM, 2017, pp. 103--117.

\end{thebibliography}

\appendix

\section{Appendix}
\label{sec:appendix}
We provide additional discussion and results considered insightful, yet tangential to the main contributions
of the proposed attack. 

\begin{table*}\label{tab:current_attacks_table}
\centering
\begin{tabular}{c|c|c|c|c|c|c|c|c}
\textbf{Attack Name} & \textbf{Attack Goal} & \textbf{Knowledge} & \textbf{Queries} & \textbf{Telephony} & \multicolumn{2}{c|}{ \textbf{Transferability} }& \textbf{Time} & \textbf{Audio Type} \\
&&&&& \textbf{Method} & \textbf{Sample} & \\
\hline

This Work & Intentional Mistranscription & Black/No & 15 & \cmark & \cmark & \cmark & ~milliseconds & Clean \\
Commander Song~\cite{yuan2018commandersong} & Hiding Signal in Audio & White & ? & \xmark & \cmark & \xmark & ? & Clean \\
Qin et al~\cite{qin2019imperceptible} & Hiding Signal in Audio & White & ? & \xmark & \xmark & \xmark & ? & Clean \\
Carlini et al.~\cite{carliniL2} & Hiding Signal in Audio & White & 1000 & \xmark & \xmark & \xmark & ? & Clean \\
M. Azalnot et al.~\cite{alzantot2018did} & Hiding Signal in Audio & Black & ? & \xmark & \cmark & \xmark & ? & Clean \\
Houdini~\cite{cisse2017houdini} & Hiding Signal in Audio & White & ? & \xmark & \xmark & \xmark & ? & Clean \\
Schonherr et al.~\cite{schonherr2018adversarial} & Hiding Signal in Audio & White & 500 & \xmark & \xmark & \xmark & minutes & Clean \\
Skill Squat~\cite{erichennenfentskill} & Hiding Signal in Audio & Black & ? & \xmark & \xmark & \xmark & ? & Clean \\
Kreuk et al.~\cite{kreuk2018fooling} & Hiding Signal in Audio & White & ? & \xmark & \xmark & \xmark & ? & Clean \\
Dolphin Attack~\cite{zhang2017dolphinattack} & Hiding Signal in near Ultrasound & Black & ? & \xmark & \cmark & \cmark & ? & Inaudible \\
HVC (1)~\cite{carlini2016hidden} & Hiding Signal in Noise & White & ? & \xmark & \xmark & \xmark & 32 hours & Noisy \\
HVC (2)~\cite{carlini2016hidden} & Hiding Signal in Noise & Black & ? & \xmark & \xmark & \xmark & ? & Noisy \\
Cocaine Noodles~\cite{vaidya2015cocaine} & Hiding Signal in Noise & Black & ? & \xmark & \xmark & \xmark & ? & Noisy \\
Abdullah et al.~\cite{biometrics} & Hiding Signal in Noise & Black & 10 & \xmark & \cmark & \cmark & ~seconds & Noisy \\
\end{tabular}

\caption{The table provides an overview of the current progress of the adversarial attacks against ASR and AVI systems. The reader can observe that our attack is different from existing works in its attack goal, adversary required knowledge, number of required queries, robustness to the telephony network, transferability, required execution time and the type of audio quality. We have designed the attack to address the specific needs of the dissident, attempting to overcome surveillance infrastructure. \\\textbf{Key:} [\xmark] Won't work or fails to demonstrate that it will. [\cmark] Will work. [?] If authors provide no information in paper. }
\label{tab:current_attacks_table}
\end{table*}

\subsection{Trivial White-Noise Attack}\label{sub:white_attack}

\subsubsection{Motivation}\label{sub:white_attack_mot}
Readers might be tempted to use trivial attacks to subvert ASR and AVI models. This includes adding white-noise to benign audio samples. However, in the following subsection, we show that any such trivial techniques will fail to achieve the dissidents' goals: fool the model whilst not impacting human interpretability of the audio sample. 

\subsubsection{Methodology and Setup}\label{sub:white_attack_meth}
We tested this white-noise method by attacking a random set of 100 audio files that contained speakers uttering a single word. We added white-noise to these samples to generate adversarial audio samples. Next, we passed both the original audio and the white-noise infused audio samples to the Google Speech API. We recorded the number of samples that were incorrectly transcribed by the API.

Next, to measure the impact on human interpretability, we used the Perceptual Evaluation of Speech Quality (PESQ) standard~\cite{rix2001perceptual}. This is a global standard used for measuring the audio quality of telephony systems. PESQ measures features such as jitter, packet loss, noise and returns a quality score between 1 (bad quality) and 5 (high quality). By using PESQ, supplants the need for user studies to measure audio quality. In our case, the PESQ score can reveal how white-noise will impact audio quality. We calculate the PESQ scores for each of the white-noise infused audio samples and calculated the average. 

\subsubsection{Results}\label{sub:white_attack_result}
Of the total audio samples we attacked, only 35\% of the samples were successfully evaded. Furthermore, average PESQ score for the samples was 1.06, which implies very low audio quality. This proves that any trivial attack will have very low attack success against models, and will have a strong negative impact on human audio interpretability. In contrast, the attack proposed in this paper has little to no impact on the human interpretability (as shown using our Amazon Turk experiments), and achieve 100\% success rate against any speech-based models.

\subsection{Impulse Perturbation Attack}

\subsubsection{Methodology}\label{sub:impulse_attack_meth}
ASR systems are trained to learn patterns using features from the training set.
It is important that training and test sets belong to the same distribution.
Otherwise, the model will have difficulty identifying patterns in the test set.
Intuitively, we may construct a simple perturbation by sampling outside of the
ASR system's training distribution, then applying it to an input to trick the
ASR system.

We extend our study of ASR system's sensitivity with this extremely simple
attack. This involves increasing the amplitude of time-samples within a single
phoneme to the maximum amplitude observed in the entire time series. This
perturbation will create a minor spike in the audio sample, known as an
\emph{impulse}. If the impulse perturbation succeeds at confusing the model, it
will highlight the high sensitivity of the model to artificial perturbations.
This will motivate further investigation of other possible attack vectors. These
can be designed to confuse the model even further, with limited to no impact on
human understandability of the attack audio sample.

The impulse perturbation described above might be able to confuse the model.
However, there are a few drawbacks to this approach. First, most popular ASR
systems are often trained on both clean and degraded audio quality. This is done
to ensure that the ASR systems perform well in noisy environments. Secondly, ASR
system's architecture is designed to ensure that even with a limited training
set, the model is able to generalize well. A better generalized model should not
be confused by such a simple perturbation. Any simple attack method will only
have limited success against ASR systems. Therefore, a further investigation of
other attack methods is necessary.

\subsubsection{Setup}\label{sub:impulse_attack_setup}
We test our simple attack on the TIMIT corpus. The TIMIT corpus contains the
timestamps of each phoneme in the audio file. First, we iteratively selected
each phoneme in a word to perturb. Then, each time sample of the target phoneme
is replaced with the largest amplitude value in the audio file. Each of the
attack audio samples is passed Over-Line to the ASR system for transcription. We
then repeat this process, but only perturb one percent of the phoneme. This
allows us to identify the relationship between the number of perturbed samples
in the phoneme and the mistranscription rate. For brevity, we tested the simple perturbation
method against a single model.

\subsubsection{Results}\label{sub:impulse_attack_result}

The simple baseline attack above is executed against the Google (Normal) model.
We observe that phonemes with one percent of their samples perturbed only had a
10\% attack success rate against the model. This number increases to 43\% when
the entire phoneme is perturbed by the impulse value. Although the attack has
limited success in this scenario, the impulses would likely fail to have an
effect against an adversarially-trained model.

\subsubsection{Simple Defense}\label{para:impulse_attack_discussion}
During our initial simple perturbation experiments, we observed that applying
\textit{impulses} to individual phonemes was easily distinguishable during
manual listening tests. Not only was the attack success rate low, but such
\textit{impulses} were reminiscent of call audio distortions and jitters that
are commonly heard over telephony networks. Using this na\"{i}ve perturbation
scheme is not ideal since a machine learning model will likely perform equally
well in distinguishing impulses as humans. Overall, adversarial training schemes
to defend against this style of attack would be trivial to implement, and do not
give an adversary sufficient probability of success under our threat model.

\begin{figure}  
  \centering
  \scalebox{0.78}{
  \begin{tabular}{l|l|c|c|c|c|c|}
  \cline{2-7}
  \multirow{2}{*}{}
    & \multirow{2}{*}{\textbf{Original Model}} 
    & \multicolumn{5}{c|}{\textbf{
      \begin{tabular}[c]{@{}c@{}}Discard
                threshold,\\ \% of max.\ DFT coeff. 
      \end{tabular}}}
    \\ \cline{3-7}  & & \textbf{2\%} & \textbf{4\%} & \textbf{8\%} & \textbf{16\%} & 
            \textbf{32\%} \\ \hline 
    \multicolumn{1}{|c|}{\textbf{\begin{tabular}[c]{@{}c@{}}Benign Test Set\\
    Accuracy (\%)\end{tabular}}} & \multicolumn{1}{c|}{85}                  &
    78.6 & 70.2 & 57.2 & 47.4 & 37.8 \\ \hline
  \end{tabular}
  }

  \vspace{0.7em}

  \includegraphics[width=1\columnwidth]{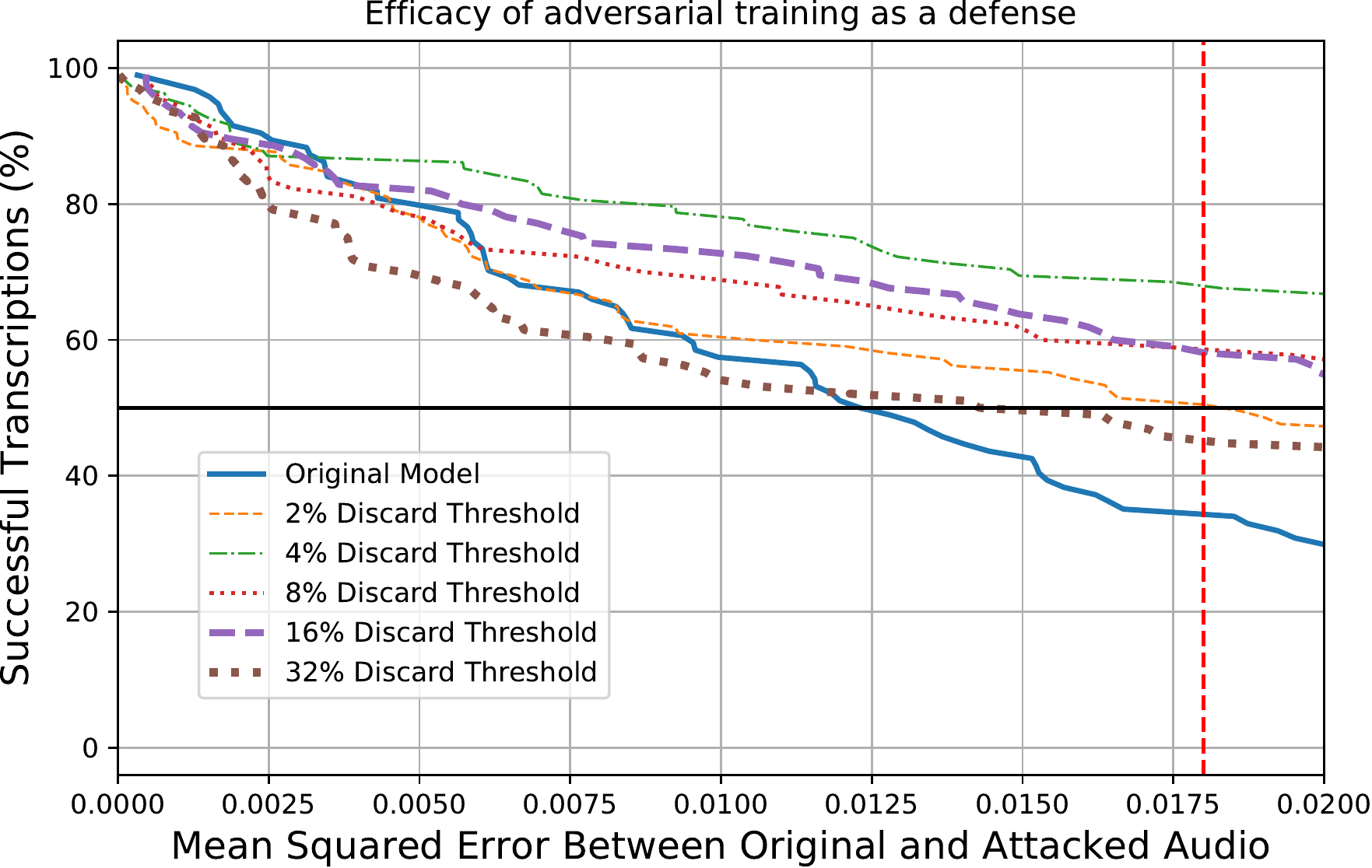}\\[10pt]
  \vspace{-1em}
  \caption{\textbf{Top:}
  The relationship between the adversarial training and the accuracy of the
  corresponding model on the benign (unmodified) test set. 
  \textbf{Bottom:}
  Transcription performance of our (small dictionary) keyword model
  when trained on audio modified to discard frequencies below various
  thresholds.  The trained models are tested on attack audio with a
  range of MSE (relative to the corresponding original audio), and the
  percentage of successful transcriptions are plotted.
  }
  \label{fig:Adv_Model_MSE}
  \label{tbl:adv_train_model_vs_acc}
  \vspace{-1.6em}
  \hrulefill
  \vspace{-1.6em}
\end{figure}

\subsection{Adversarial Training as a Defense:} \label{sub:adv_training}
One technique that has shown promise in defending computer vision models is
adversarial training~\cite{madry2017towards}. However, this approach has not
seen much success in defending speech and voice identification
models~\cite{biometrics}. To test this technique against our attack, we trained
six keyword recognition models~\cite{tf_speech_rec}. For each model, we
generated adversarial data using the method described
Section~\ref{sub:word_level_perturbations}. The threshold was determined as a
percentage of the maximum spectral magnitude (i.e., $\max_k |f_k|$); in
particular, we considered 2\%, 4\%, 8\%, 16\%, and 32\% for each of our models,
shown in Figure~\ref{fig:Adv_Model_MSE}. For example, if the threshold is 4\%,
we only retain the $f_k$ whose magnitude is greater than 4\% of the maximum
magnitude. Each of the models was trained to detect 10 keywords.

Next, we evaluated each model by randomly selecting 20 samples per keyword.
Figure~\ref{fig:Adv_Model_MSE} displays the results of our experimentation.
Figure~\ref{fig:Adv_Model_MSE} (\textbf{Top}) shows the accuracy of each model on
the benign data set, while Figure~\ref{fig:Adv_Model_MSE} (\textbf{Bottom}) shows
the transcription success at various levels of acoustic distortion (relative to
the original audio) introduced by our attack. The red dotted line represents the
limit of human comprehension as defined earlier in
Section~\ref{sub:word_level_perturbations_result}.

There are a few important trends to note. First, models trained with higher
threshold values have lower accuracy on normal audio samples, shown in
Figure~\ref{fig:Adv_Model_MSE} (\textbf{Top}). This result is expected, as lower
accuracy is an artifact of adversarial training~\cite{tsipras2018robustness,
dohmatob2018limitations, kurakin2016adversarial}. 
Second, as MSE increases, the transcription success rate decreases, shown in
Figure~\ref{fig:Adv_Model_MSE} (\textbf{Bottom}). The more samples that have
lower MSE behind the red dotted GSM line, the more sensitive the model is to our
attack.  Lastly, adversarial training does decrease model sensitivity to our
attack, relative to the baseline model. Intuitively, this implies a relationship
between the amount of adversarial training and the minimum amount of distortion
caused by our attack. 

We caution against taking Figure~\ref{fig:Adv_Model_MSE} as strong evidence
that adversarial training is a defense against our attack.  First, model
sensitivity is measured in the number of samples on the left the red GSM line in
Figure~\ref{fig:Adv_Model_MSE}. We used the GSM line as the dividing point
between what is and is not comprehensible by human listeners, as discussed in
Section~\ref{sub:asr_attacks}. Hence we consider attack audio with MSE values to
the right of this line to be failed attack samples. Yet the GSM line should be viewed as
a \emph{conservative} minimum for human comprehension.
This is important because, for a given model, our attack may produce
many audio samples whose MSE is to the right of the GSM line.  Yet,
this \emph{does not} imply that the model is necessarily ``robust'' against
our attack.  In particular, some high MSE attack samples \emph{may still be understandable
by humans} while inducing errors in the model.
Second, our experimental setup was designed only to support a
preliminary investigation of adversarial training as a defense.
It would be incorrect to extrapolate any trends from such a simple
experiment. A broader and more comprehensive examination should
consider (in detail) the effects of different
model's hyper-parameters, and employ a much larger number of audio
samples.  We leave such a study for future work.

\subsection{Supplemental Results}\label{app:results}

\begin{figure}[t!]
  \centering
  \includegraphics[width=0.8\columnwidth, height=6cm]{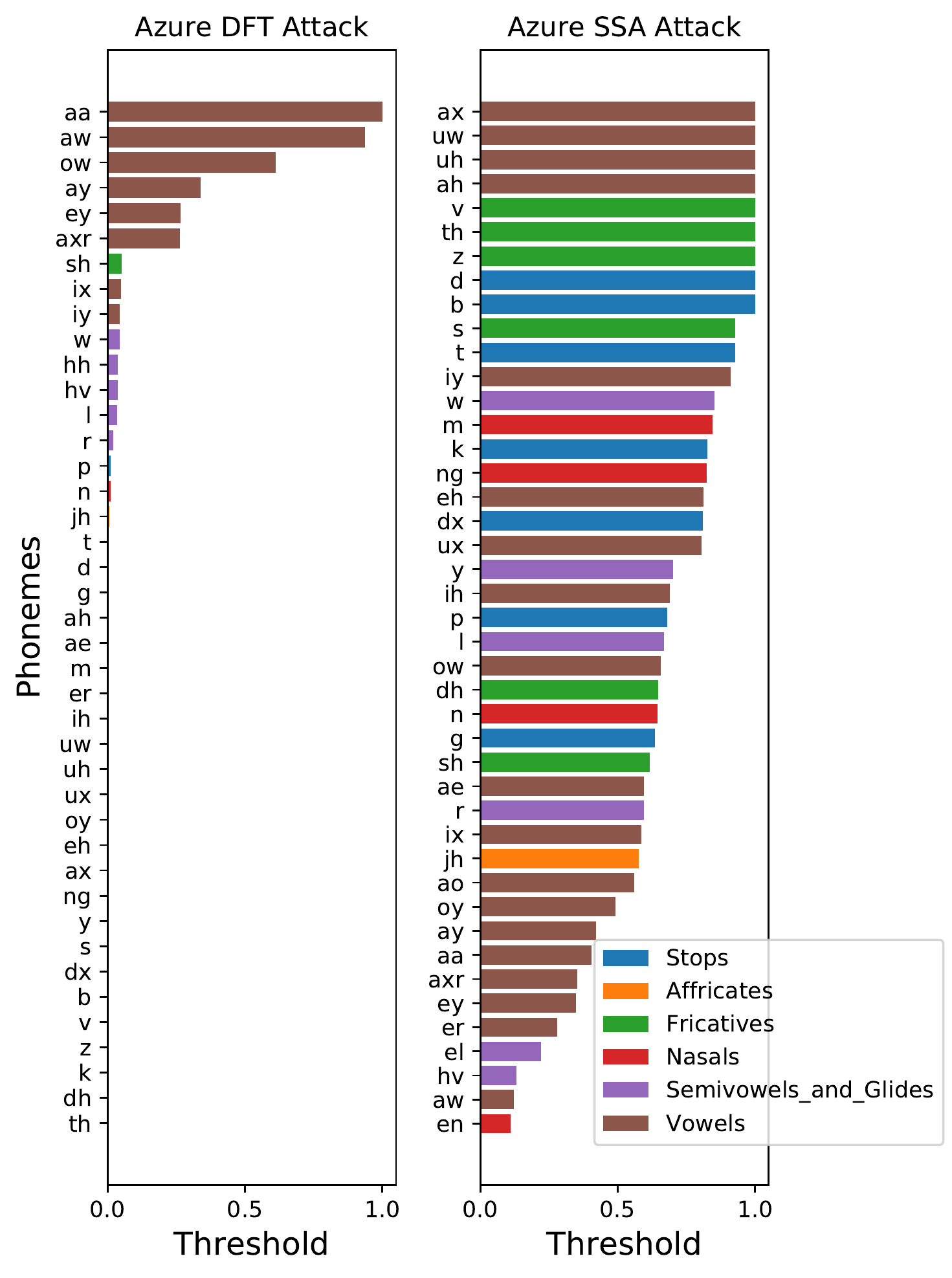}
  \caption{The graph above shows the attacks against a voice identification model.
  It shows the minimum threshold value required, when perturbing only a single
  phoneme, to successfully force the model to mis-classify the speaker. We can
  observe that in general, SSA attacks require much higher thresholds to
  successfully fool the model, in comparison to the DFT attack. }
  \label{fig:Azure_phoneme_comp_threshold}
  \vspace{-1.6em}
  \hrulefill
  \vspace{-1.6em}
\end{figure}

\begin{figure*} 
  \centering

    \subfloat[t][DFT Attack Results.]{\includegraphics[width=0.8\columnwidth]{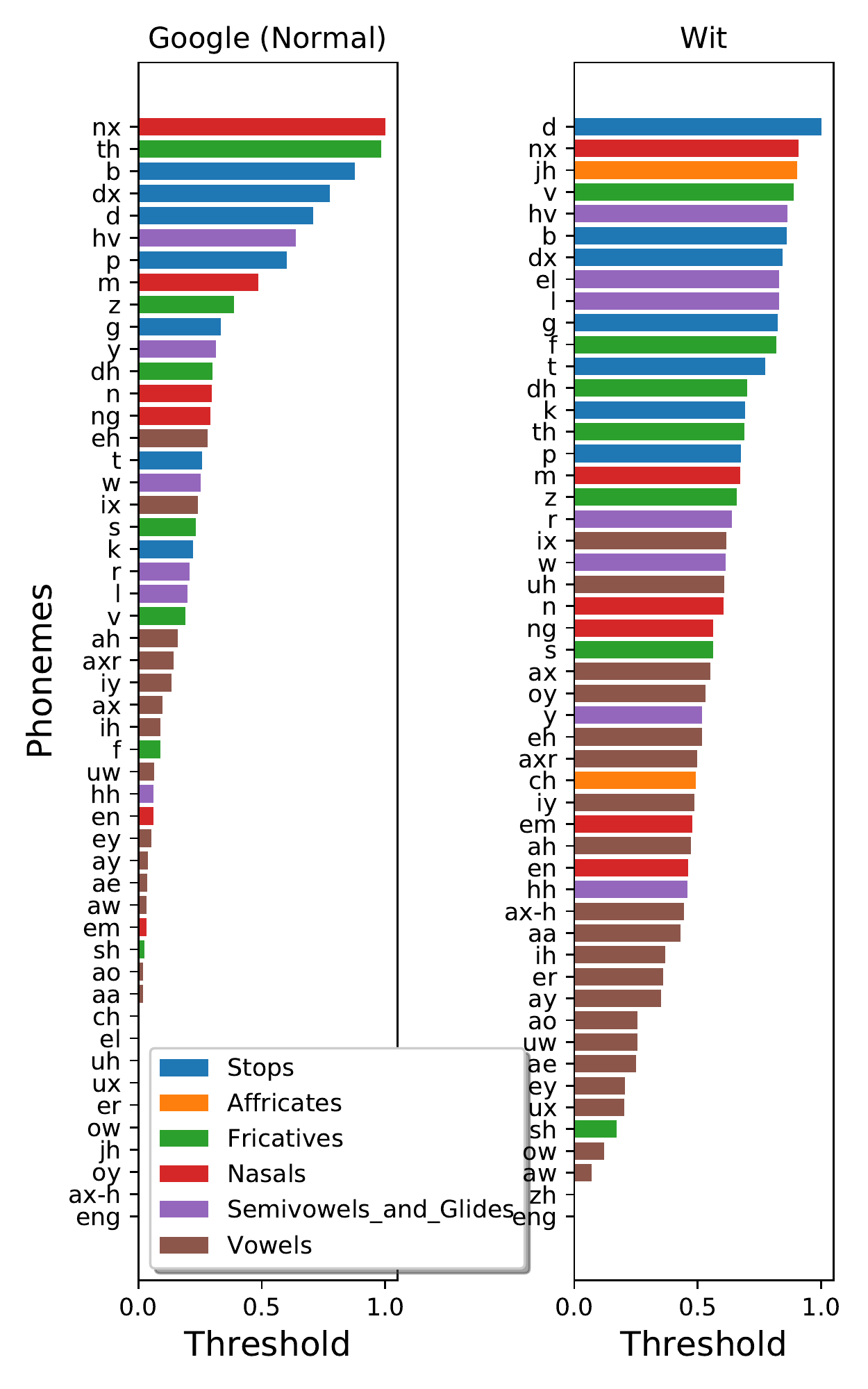}}
  \qquad
    \subfloat[t][SSA Attack Results]{\includegraphics[width=0.8\columnwidth]{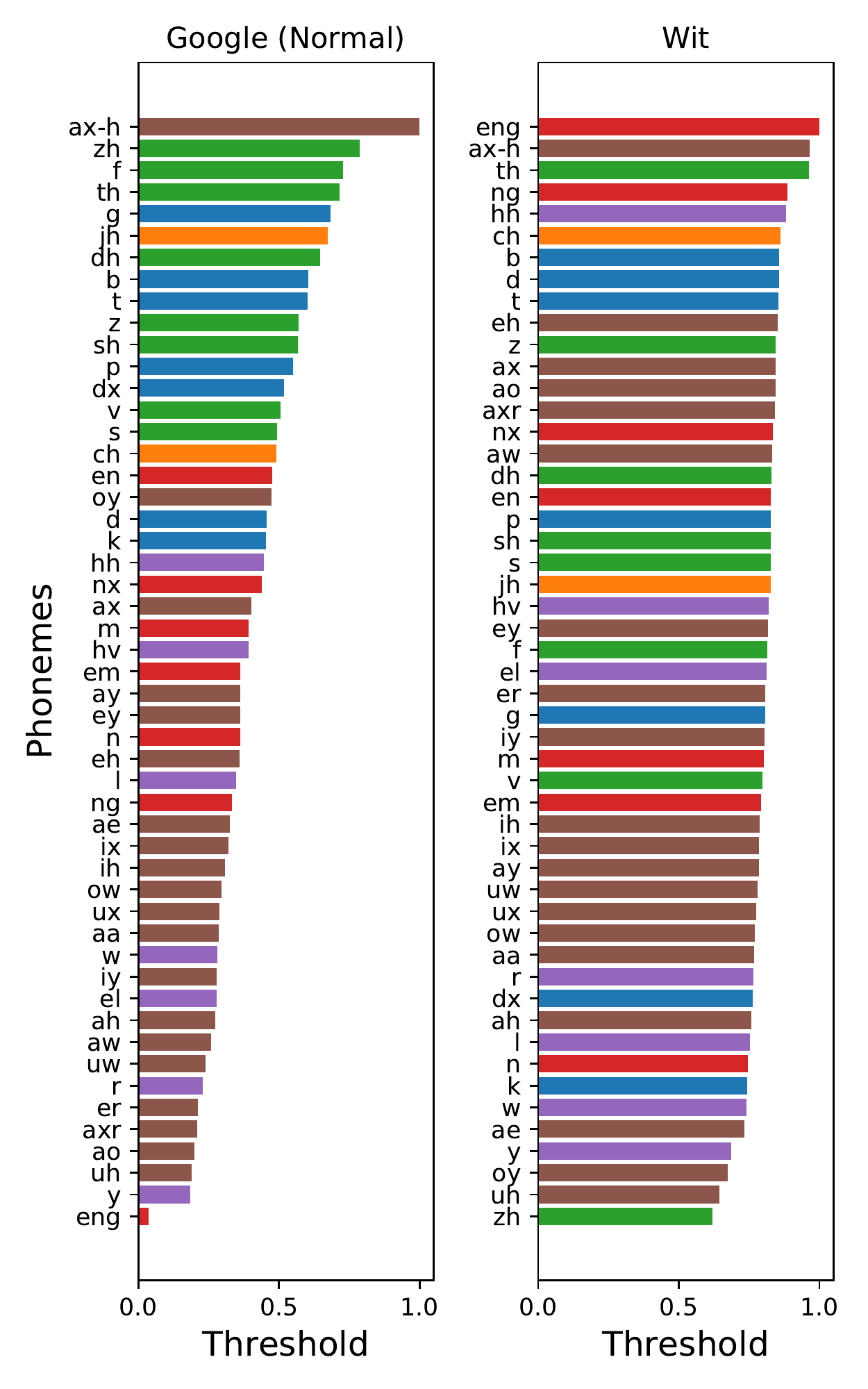}}
  \caption{Phoneme vulnerability for a selected models, Wit and Google (Normal), using our phoneme-level attacks. Lower threshold corresponds to lower distortion required for an attack success. The DFT attack exhibits a pattern of targeting the vowels (Brown) more effectively than other phonemes. In contrast, the SSA attack does not display any such consistent behavior.}

 \label{fig:phoneme_comp}
  \vspace{-1.6em}
  \hrulefill
  \vspace{-1.6em}
\end{figure*}

\end{document}